\documentclass[twocolumn,twocolappendix]{aastex7}
\usepackage{graphicx}
\usepackage{amsmath}
\usepackage{bm}
\usepackage[utf8]{inputenc}
\usepackage[T1]{fontenc}
\newcommand{\Msun}{M_{\sun}}
\newcommand{\Zsun}{Z_{\sun}}
\newcommand{\bl}[1]{\mbox{\boldmath$ #1 $}}

\begin{document}

\title{Simulations of multiple dust ring formation in a subsolar-metallicity protoplanetary disk}

\author{Ryoki Matsukoba}
\affiliation{National Institute of Technology, Kochi College, 200-1 Monobe, Nankoku, Kochi 783-8508, Japan}
\affiliation{Center for Computational Sciences, University of Tsukuba, Ten-nodai, 1-1-1 Tsukuba, Ibaraki 305-8577, Japan}
\affiliation{Department of Physics, Graduate School of Science, Kyoto University, Sakyo, Kyoto 606-8502, Japan}
\email{matsukoba@gm.kochi-ct.jp}

\author{Eduard I. Vorobyov}
\affiliation{Department of Astrophysics, University of Vienna, Tuerkenschanzstrasse 17, 1180, Vienna, Austria}
\email{eduard.vorobiev@univie.ac.at}
\affiliation{Ural Federal University, 19 Mira Str., 620002 Ekaterinburg, Russia}

\author{Takashi Hosokawa}
\affiliation{Department of Physics, Graduate School of Science, Kyoto University, Sakyo, Kyoto 606-8502, Japan}
\email{hosokawa@tap.scphys.kyoto-u.ac.jp}

\begin{abstract}
Super-Earths exist around subsolar-metallicity host stars with a frequency comparable to that around solar-metallicity stars, suggesting efficient assembly of dust grains even in metal-deficient environments. In this study, we propose a pathway for the formation of multiple dust rings that will promote planetesimal formation in a subsolar-metallicity disk. We investigate the long-term evolution of a circumstellar disk with 0.1\:$\Zsun$ over 750\:kyr from its formation stage using two-dimensional thin-disk hydrodynamic simulations. The motion of dust grains is solved separately from the gas, incorporating dust growth and self-consistent radial drift. The disk is initially gravitationally unstable and undergoes intense fragmentation. By 300\:kyr, it tends toward a stable state, leaving a single gravitationally bound clump. This clump generates tightly wound spiral arms through its orbital motion. After the clump dissipates at $\sim$410\:kyr, the spiral arms transition into axisymmetric substructures under the influence of viscosity. These axisymmetric substructures create local gas pressure bumps that halt the inward radial drift of dust grains, resulting in the formation of multiple-ring-shaped dust distributions. We observe several rings within $\simeq 200$\:au of the central star, with separations between them on the order of $\sim 10$\:au, and dust surface density contrasts with inter-ring gaps by factors of $\sim 10-100$. We also demonstrate that turbulent viscosities at observationally suggested levels are essential for converting spiral arms into axisymmetric substructures. We speculate that the physical conditions in the dust rings may be conducive to the development of streaming instability and planetesimal formation.
\end{abstract}


%%%%%%%%%%%%%%%%%%%%%%%%%%%%%%%%%%%%%%%%%%%%%
%%%%%%%%%%%%%%%%% BODY OF PAPER %%%%%%%%%%%%%%%%%%

%%%%%%%%%%%%%%%%%%%%%%%%%%%%%%%%%%%%%%%%%%%%
%%%%%%%%%%%%%%%%%%%%%%%%%%%%%%%%%%%%%%%%%%%%
%%% SECTION 1 %%%%
\section{Introduction}
\label{Sec:Intro}

%---------------------------------------------------%

The number of discovered exoplanets has exceeded 5,000 (the NASA Exoplanet Archive; \citealt{Akeson2013}), providing valuable insight into their statistical properties. The occurrence rate of planets depends on the metallicity of their host stars, but this dependence varies by planet type. Giant gas and sub-Neptune planets are more common around stars with higher metallicity \citep[e.g.][]{Santos2004, Fischer2005, Petigura2018, Fulton2021}. 
In contrast, the occurrence rate of super-Earth planets appears to show little \citep{Wang2015} to no correlation with metallicity \citep{Mulders2016, Kutra2021}. These relationships with metallicity imply variations in the formation processes across different planet types and suggest that the formation of super-Earth precursors is less affected by environmental conditions.

%---------------------------------------------------%

The properties of protoplanetary disks with subsolar metallicity differ from those with solar metallicity \citep[e.g.][]{Guadarrama2022,Gehrig2023}. A subsolar-metallicity disk contains less dust than a solar-metallicity disk, and the lower dust abundance slows dust growth \citep{Birnstiel2016}. Since dust drift motion depends on dust size \citep{Weidenschilling1977}, this affects the spatial distribution of dust within the disk. Additionally, the amount of dust influences the disk's thermal evolution. Analytical and numerical studies have shown that disks with less dust are more gravitationally unstable due to differences in thermal evolution \citep{Tanaka2014, Bate2014, Machida2015, Bate2019, Matsukoba2022}. In the solar-metallicity case, both the accretion envelope and the disk are efficiently cooled by dust thermal emission. At lower metallicities, however, dust cooling becomes ineffective in low-density regions such as the envelope. As a result, the envelope becomes progressively warmer compared to the disk as the metallicity decreases. Because the infall rate onto the disk depends on the envelope temperature, the mass supply from the envelope to the disk becomes higher than the accretion rate from the disk onto the protostar. This imbalance can also occur temporarily in young solar-metallicity disks \citep{Vorobyov2005}, its magnitude increases with lower metallicity, causing subsolar-metallicity disks to accumulate more mass and become denser than their solar-metallicity counterparts \citep[e.g.][]{Kratter2010,Tanaka2014}. Such massive and compact disks are therefore more susceptible to fragmentation, potentially leading to the formation of multiple stellar systems.

%---------------------------------------------------%

For planet formation, micron-sized dust grains must grow into kilometer-sized planetesimals, which then further develop into protoplanets. In a disk with subsolar metallicity, not only is dust growth slow, but the lower dust content also makes planet formation more challenging. The decrease in the occurrence rates of giant gas and sub-Neptune planets around subsolar-metallicity stars aligns with these challenges. However, the weak or absent metallicity dependence of the occurrence rate of super-Earths is counterintuitive, indicating that the formation of planetesimals and protoplanets capable of producing rocky planets of such sizes is still feasible even in subsolar-metallicity disks. Recent and timely theoretical studies have explored the potential for planetesimal formation in very metal-poor environments \citep{Eriksson2025, Vorobyov2025}. In particular, \citet{Eriksson2025} show that efficient dust trapping by vortices can concentrate grains, enabling planetesimal formation above a metallicity threshold of $\sim 0.04\:\Zsun$. A comparable threshold was obtained by \citet{Vorobyov2025}, who considered dead zones in low-metallicity disks with radially varying gravitational instability.

%---------------------------------------------------%

Multiple dust rings and gaps have been observed in many protoplanetary disks \citep{ALMA2015, Andrews2016, Andrews2018, Clarke2018, Huang2018a, Huang2018b, Isella2018, Facchini2020, Huang2020}. Within the ring, dust is concentrated, promoting collisions and accelerating dust growth. Therefore, dust rings are considered promising sites for planet formation. If dust rings form in a subsolar-metallicity disk, the local dust-to-gas mass ratio is enhanced, potentially mitigating the challenges of planet formation in metal-deficient environments.

%---------------------------------------------------%

The formation mechanism of dust rings and gaps is not yet fully understood. Proposed scenarios include disk-planet interaction \citep[e.g.][]{Zhu2014, Dong2017, Bae2018a, Bae2018b, Meru2019, Weber2019, Wafflard-Fernandez2020}, magnetohydrodynamic effects \citep[e.g.][]{Varniere2006, Flock2015, Suriano2018, Riols2019}, ice lines \citep[e.g.][]{Zhang2015, Okuzumi2016, Pinilla2017}, and secular gravitational instability \citep[e.g.][]{Youdin2011, Takahashi2014, Tominaga2018, Tominaga2020, Tominaga2023}. A common feature of these mechanisms is the formation of axisymmetric, locally dense gas pressure bumps. In such regions, dust grains are trapped at these pressure bumps due to their drift motion.

%---------------------------------------------------%

The formation of dust rings in solar metallicity disks has been a key topic in both observational and theoretical studies. However, such a process in the lower-metallicity environments and the implications for planet formation have received comparatively little attention. In this study, we investigate the formation of multiple dust rings and gaps in a disk with one-tenth of solar metallicity using a two-dimensional hydrodynamic simulation. Our simulation begins with the gravitational collapse of a pre-stellar core and follows the formation and evolution of the disk. We solve the evolution of the gas disk while consistently tracking the size growth of dust grains from micron sizes and their drift motion. This simulation follows the long-term evolution up to 750\:kyr after disk formation. Thanks to this long-term approach, we identify a new mechanism for the formation of multiple dust rings and gaps. We note that we previously performed long-term simulations of a subsolar-metallicity disk \citep{VorobyovElbakyan2020b, Matsukoba2023}; however, these did not take dust-size growth into account. We show that, in early phases, non-axisymmetric spiral arms emerge in the self-gravitationally unstable disk. As the disk stabilizes, these spiral arms transition into axisymmetric structures, where dust accumulates. This mechanism is favorable in metal-deficient environments, as disks with lower metallicity tend to be more gravitationally unstable. 

%---------------------------------------------------%

This paper is organized as follows: We describe our simulation method and setup in Section\:\ref{Sec:Method}. We present the simulation results and explain the disk evolution and formation of multiple dust rings and gaps in Section\:\ref{Sec:DiskEvolution}. We then examine the results of numerical experiments in Section\:\ref{Sec:NumericalExperiments}. Finally, the summary and discussion are provided in Section\:\ref{Sec:Summary}.

%%%%%%%%%%%%%%%%%%%%%%%%%%%%%%%%%%%%%%%%%%%%
%%%%%%%%%%%%%%%%%%%%%%%%%%%%%%%%%%%%%%%%%%%%
%%% SECTION 2 %%%%
\section{Method}
\label{Sec:Method}

%---------------------------------------------------%

We simulate gas and dust dynamics in a protoplanetary disk with one-tenth of solar metallicity using the Formation and Evolution of Stars and Disks (FEOSAD) code presented in \citet{VorobyovAkimkin2018} and modified to include the backreaction of dust on gas in \citet{VorobyovElbakyan2020}. FEOSAD solves the equations of hydrodynamics in the thin-disk limit for a gas--dust system. The numerical simulation starts from the gravitational collapse of a flattened pre-stellar cloud, and the protoplanetary disk is formed self-consistently as a result of angular momentum conservation of the contracting cloud. In this work, we apply FEOSAD to non-solar metallicity environments and the usual energy balance equation is modified to include thermal processes that may be important at subsolar metallicity as described in \citet{VorobyovMatsukoba2020}. 

%---------------------------------------------------%

The integration of the hydrodynamic equations in the polar coordinates ($r$, \:$\phi$) is carried out using a finite-volume method with a time-explicit solution procedure similar in methodology to the ZEUS code \citep{Stone1992}. The advection of gas and dust is treated using the third-order-accurate piecewise-parabolic interpolation scheme of \citet{Colella1984}, which is characterized by very low numerical diffusion. Owing to consistent transport \citep{Norman1980}, our numerical scheme globally conserves angular momentum within machine precision. When the disk achieves a steady-state condition, the code can hold the disk in equilibrium for hundreds of orbital periods \citep{Vorobyov2006}. The stellar mass grows according to the mass accretion rate through the inner computational boundary and the properties of the protostar are calculated using the stellar evolution tracks obtained with the STELLAR code for one-tenth of solar metallicity \citep{Yorke2008, Hosokawa2009, Hosokawa2013}. 

%---------------------------------------------------%

The central 10\:au are cut out and replaced with a sink cell. We adopt a free inflow--outflow boundary condition, which allows material to flow both from the computational domain to the sink cell and vice versa. The details of the boundary condition are discussed in \cite{VorobyovAkimkin2018}. The inflow--outflow boundary condition performs better than the standard outflow boundary condition, because it allows for a compensating flow from the sink to the computational domain when oscillatory motions are present at the boundary interface \citep{Zhu2012}. Since we are interested in the effects occurring far away from the inner computational boundary, we do not expect any notable influence of the adopted boundary conditions on our results (see Appendix\:\ref{Sec:Boundary}). The entire computational domain extends from 10\:au to 8800\:au. The radius of the computational domain is sufficiently large compared to the disk's radial scale, ensuring that the outer computational boundary does not affect the evolution of the disk.  

%---------------------------------------------------%

Below, we provide the basic equations and initial conditions. More details can be found in the aforementioned articles.

%%%%%%%%%%%%%%%%%%%%%%%%%%%%%%%%%%%%%%%%%%%%
%%% Section 2.1 %%%%
\subsection{Basic equations for the gaseous component}

The system of equations for the gaseous component consists of the continuity equation, equations describing the gas dynamics, and the energy balance equation. The dynamics of gas is governed by the stellar and disk gravity, turbulent viscosity, and friction between gas and dust. The pertinent equations in the thin-disk limit are as follows.
\begin{equation}
\label{eq:cont}
    \frac{{\partial \Sigma_{\rm g} }}{{\partial t}} 
    + \nabla_{\rm p} \cdot \left( \Sigma_{\rm g} {\bl v}_{\rm p} \right) = 0,  
\end{equation}
\begin{eqnarray}
\label{eq:mom}
    \frac{\partial}{\partial t} \left( \Sigma_{\rm g} {\bl v}_{\rm p} \right) 
    & +& \left[\nabla \cdot \left( \Sigma_{\rm g} {\bl v}_{\rm p} \otimes {\bl v}_{\rm p} \right)\right]_{\rm p} \nonumber \\ 
    & =& - \nabla_{\rm p} {\cal P}  
    + \Sigma_{\rm g} \, {\bl g}_{\rm p}
    + \left(\nabla \cdot \mathbf{\Pi}\right)_{\rm p} 
    - \Sigma_{\rm d,gr} {\bl f}_{\rm p},
\end{eqnarray}
where the planar components ($r$,\:$\phi$) are denoted by the subscript ${\rm p}$, $\Sigma_{\rm g}$ is the gas surface density, ${\bl v}_{\rm p}=v_r\hat{{\bl r}}+v_\phi \hat{{\bl \phi}}$ is the gas velocity in the disk plane, $\cal{P}$ is the pressure, integrated in the vertical direction using the ideal equation of state ${\cal P}=(\gamma-1) e$ with the internal energy $e$, and ${\bl f}_{\rm p}$ is the drag force per unit mass between gas and dust. The gravitational acceleration in the disk plane ${\bl g}_{\rm p}$ takes into account gas and dust self-gravity and the gravity of the central star when it is formed. The star is introduced when the gas mass in the sink cell (see its definition below) exceeds 0.02~$M_\odot$, which is just a few times greater than the mass of the protostellar seed \citep{MasunagaInutsuka2000}. The gravitational potential of the gas and dust components is calculated using the convolution method described in detail in \citet{Binney1987} and \citet{VorobyovSkliarevskii2024}. This method is free from the smoothing-length assumption, which is often used to avoid the singularity in the gravitational potential calculations \citep[e.g.,][]{Baruteau2008, Hure2009, Mueller2012}. We note that the smoothing-length method can achieve a slightly better accuracy in computing the gravitational potential if  the proper choice of the smoothing factor is applied. However, our approach is more stable because there is no universal formula to find the best-fit smoothing length. The necessary tests and details of the advantages and disadvantages of our approach to compute the gravitational potential are discussed in the Appendix of \cite{VorobyovSkliarevskii2024}. 

%---------------------------------------------------%

To compute the viscous stress tensor ${\bl \Pi}$, we parameterize the kinematic viscosity using the usual $\alpha$-parameter approach with the value of $\alpha$ set equal to $10^{-4}$ throughout the entire domain. This low $\alpha$ parameter is an assumption of our model and is supported by observational evidence suggesting low $\alpha$ parameters in solar metallicity environments \citep[e.g.,][]{Flaherty2015, Pinte2016, Flaherty2020, Trapman2020}. As was shown in \citet{Vorobyov2009}, the dynamics of gravitationally unstable disks for such a low $\alpha$-value is almost totally dominated by disk self-gravity and associated gravitational torques. The reason why we still keep turbulent viscosity in the equations and do not set it simply to zero is that it plays an important role in dust growth (see Section\:\ref{Sect:dust}). Our code also uses artificial viscosity to smooth shock fronts over two grid cells, but its effect on disk evolution is negligible compared to turbulent viscosity caused by gravitational instability and represented by the $\alpha$ parameter, as shown in \cite{Vorobyov2007}.

%---------------------------------------------------%

The internal energy balance equation is written as
\begin{equation}
    \frac{\partial e}{\partial t} +\nabla_{\rm p} \cdot \left( e {\bl v}_{\rm p} \right) 
    = -{\cal P} \left(\nabla_{\rm p} \cdot {\bl v}_{\rm p}\right) - \Lambda 
    + \left(\nabla {\bl v}\right)_{\rm pp^\prime}:\Pi_{\rm pp^\prime}, 
\label{eq:energ}
\end{equation}
where $\Lambda$ encompasses the cooling/heating processes that are pertinent to protoplanetary disks at one-tenth of solar metallicity. In particular, we considered the following processes: the continuum emissions of gas and dust, molecular line emissions of H$_2$ and HD, fine-structure line emissions of O\:{\sc I} ($63\:\mu{\rm m}$) and C\:{\sc II} ($158\:\mu{\rm m}$), and chemical cooling/heating associated with H ionization/recombination and  H$_2$ dissociation/formation. When we calculate the rates of these emission processes, the opacities of gas and dust are taken into account using the opacity tables from \cite{Semenov2003}. The dust opacity in each grid cell is scaled according to a local dust-to-gas mass ratio. In addition, we solve the non-equilibrium chemical network for eight species, H, H$_2$, H$^+$, H$^-$, D, HD, D$^+$, and e$^-$, with 27 reactions. The evolution of the chemical components affects the rates due to molecular line emissions and chemical cooling/heating. The dust temperature is calculated from the energy balance of dust grains as a result of the thermal emission, absorption, and collision with gas. This approach permits decoupling of the gas and dust temperatures in the low-density or high-temperature regime or at subsolar metallicity, the latter being the case in our study. Therefore, our scheme allows for a more accurate description of subsolar-metallicity disks than the schemes that assume similar gas and dust temperatures and simply scale down the dust and gas opacities. More details on the thermal scheme applied in this work can be found in \citet{VorobyovMatsukoba2020}.

%%%%%%%%%%%%%%%%%%%%%%%%%%%%%%%%%%%%%%%%%%%%
%%% Section 2.2 %%%%
\subsection{Basic equations for the dust component}
\label{Sect:dust}

The dust component in our model is divided into two populations: (i) small dust, which are grains with a size\footnote{By the size of dust grains we refer to its radius.} between $a_{\rm min}=5\times 10^{-3} \ \mu \rm m$  and $a_{*} = 1 \ \mu \rm m$ and (ii) grown dust ranging in size from $a_{*}$ to a maximum value $a_{\rm max}$, which is variable in space and time. Initially, all dust in a collapsing pre-stellar cloud is in the form of small dust grains. Small dust  can grow and turn into grown dust as the disk forms and evolves. Dust in both populations is distributed over size according to a simple power law: 
\begin{equation}
    N(a) = C \cdot a^{ - q},
\label{eq:dustdistlaw}
\end{equation} 
where $N(a)$ is the number of dust particles per unit dust size, $C$ is a normalization constant, and $q = 3.5$. We note that the power index $q$ is kept constant during the considered disk evolution period for simplicity.

%---------------------------------------------------%

We solve the continuity equations separately for the grown and small dust ensembles. However, the momentum equation is solved only for the grown dust, because small dust is assumed to be dynamically linked to the gas. The system of hydrodynamics equations for the two-population dust ensemble in the zero-pressure limit is written as:
\begin{equation}
\label{contDsmall}
    \frac{{\partial \Sigma_{\rm d,sm} }}{{\partial t}} 
    + \nabla_{\rm p} \cdot \left( \Sigma_{\rm d,sm} {\bl v}_{\rm p} \right) 
    = - S(a_{\rm max}),  
\end{equation}
\begin{equation}
\label{contDlarge}
    \frac{{\partial \Sigma_{\rm d,gr} }} {{\partial t}} 
    + \nabla_{\rm p} \cdot \left( \Sigma_{\rm d,gr} {\bl u}_{\rm p} \right) 
    = S(a_{\rm max}),
\end{equation}
\begin{eqnarray}
\label{eq:momDlarge}
    \frac{\partial}{\partial t} \left( \Sigma_{\rm d,gr} {\bl u}_{\rm p} \right) 
    &+& \left[\nabla \cdot \left( \Sigma_{\rm d,gr} {\bl u}_{\rm p} \otimes {\bl u}_{\rm p} \right)\right]_{\rm p} \nonumber \\ 
    &=& \Sigma_{\rm d,gr} \, {\bl g}_{\rm p} 
    + \Sigma_{\rm d,gr} \, {\bl f}_{\rm p} 
    + S(a_{\rm max}) {\bl v}_{\rm p},
\end{eqnarray}
where $\Sigma_{\rm d,sm}$ and $\Sigma_{\rm d,gr}$ are the surface densities of small and grown dust, respectively, and ${\bl u_{\rm p}}$ are the planar components of the grown dust velocity.

%---------------------------------------------------%

The grown dust dynamics is sensitive to the properties of surrounding gas. The drag force (per unit mass) links dust with gas and can be written according to \citet{Weidenschilling1977} as:
\begin{equation}
    {{\bl f}_{\rm p}} 
    = \dfrac{1} {2 m_{\rm d}} C_{\rm D} \, \sigma \rho_{\rm g} ({{\bl v_{\rm p}}} - {{\bl u_{\rm p}}}) |{{\bl v_{\rm p}}} - {{\bl u_{\rm p}}}|,
\label{eq:friction}
\end{equation}
where $\sigma$ is the dust grain cross section, $\rho_{\rm g}$ the volume density of gas, $m_{\rm d}$ the mass of a dust grain, and $C_{\rm D}$ the dimensionless friction parameter. The functional expression for the latter is taken from \citet{Henderson1976} and is described in detail in   \citet{Stoyanovskaya2020} and \citet{Vorobyov2023}. The use of the Henderson friction coefficient allows us to treat the drag force in Epstein and Stokes regimes self-consistently, reproducing the effects of the sonic boom and avoiding discontinuities that are present in the more common drag coefficient proposed by \citet{Weidenschilling1977}. To account for the back-reaction of grown dust on gas, the term $\Sigma_{\rm d,gr} f_{\rm p}$ is symmetrically included in both the gas and dust momentum equations. We use the maximum size of dust grains $a_{\rm max}$ when calculating the value of $m_{\rm d}$ in Equation\:(\ref{eq:friction}). This calculation of $m_{\rm d}$ may slightly underestimate the friction force $|{\bl f}_{\rm p}|$.   Indeed, for our choice of $q=3.5$, we found that about 55\%--60\% of the total grown dust mass is located in the dust size bin [$0.2 a_{\rm max}:a_{\rm max}$] when $a_{\rm max} > 30\, a_\ast$, and the corresponding mass fraction is 60\%--65\% when $5 a_\ast< a_{\rm max} \le 30 a_\ast$.  Since most dust mass carriers are localized in the [$0.2 a_{\rm max}:a_{\rm max}$] bin, a more accurate calculation of $|{\bl f}_{\rm p}|$ involves weighting $m_{\rm d}$ and $\sigma$ in Equation~(\ref{eq:friction}) over this dust size bin, which we leave for future work.

%---------------------------------------------------%

The term $S(a_{\rm max})$ that enters the equations for the dust component is the conversion rate between small and grown dust populations. We assume that the distribution of dust particles over size follows the form given by Equation\:\eqref{eq:dustdistlaw} for both small and grown populations. Furthermore, the distribution is assumed to be continuous at $a_{*}$. Our scheme is constructed so as to preserve continuity at $a_{*}$ by writing the conversion rate of small to grown dust in the following form:
\begin{equation}
    S(a_{\rm max}) = - \dfrac{\Delta \Sigma_{\rm d,sm}} {\Delta t},
\end{equation}
where
\begin{eqnarray}
\label{final}
    &\Delta&\Sigma_{\mathrm{d,sm}} = \Sigma_{\mathrm{d,sm}}^{n+1}- \Sigma_{\mathrm{d,sm}}^{n} \nonumber \\ &=&
    \frac
    {
    \Sigma_{\rm d,gr}^n \int_{a_{\rm min}}^{a_*} a^{3-q}{\rm d}a - 
    \Sigma_{\rm d,sm}^n \int_{a_*}^{a_{\mathrm{max}}^{\rm n+1}} a^{3-q}{\rm d}a
    }
    {
    \int_{a_{\rm min}}^{a_{\mathrm{max}}^{n+1}} a^{3-q}{\rm d}a
    },
\end{eqnarray}
where indices $n$ and $n+1$ denote the current and next hydrodynamic steps of integration, respectively, and $\Delta t$ is the hydrodynamic time step. The adopted scheme effectively assumes that dust growth smooths out any discontinuity in the dust size distribution at $a_\ast$ that may appear due to differential drift of small and grown dust populations. A more detailed description of the scheme is presented in\:\citet{Molyarova2021} and \citet{VorobyovSkliarevskii2022}. 

%---------------------------------------------------%

The conversion rate $S(a_{\rm max})$ depends only on the local maximal size of dust $a_{\rm max}$, since the values of $a_{\rm min}$ and $a_\ast$ are fixed in our model. At the beginning of the cloud core collapse, all grains are in the form of small dust, namely, $a_{\rm max} = 1.0$\:$\mu$m. During the disk formation and evolution epoch the maximal size of dust particles usually increases. The change in $a_{\rm max}$ within a particular numerical cell in our model occurs due to collisional growth or via advection of dust through the cell. The equation describing the dynamical evolution of $a_{\rm max}$ is as follows:
\begin{equation}
    \frac{\partial a_{\rm max}}{\partial t} 
    + ({\bl u}_{p} \cdot \nabla_p ) a_{\rm max} = \cal{D},
\label{eq:dustA}
\end{equation}
where the rate of dust growth due to collisions and coagulation is computed in the monodisperse approximation \citep{Birnstiel2012}
\begin{equation}
    \cal{D} = \frac{\rho_{\rm d} \mathit{u}_{\rm rel}}{\rho_{\rm s}}.
\end{equation}
This rate includes the total volume density of dust $\rho_{\rm d}$, the dust material density $\rho_{\rm s} = 3.0$\:g\:cm$^{-3}$, and the relative velocity of particle-to-particle collisions defined as $\mathit{u}_{\rm rel} = (\mathit{u}_{\rm th}^2 + \mathit{u}_{\rm turb}^2)^{1/2}$, where $\mathit{u}_{\rm th}$ and $\mathit{u}_{\rm turb}$ account for the Brownian and turbulence-induced local motion, respectively. In particular, the dust-to-dust collision velocity owing to turbulence is computed following the model of turbulent eddies proposed in \citet{Ormel2007} 
\begin{equation}
    u_{\rm{turb}} = \sqrt{{\frac{3 \alpha}{\mathrm{St}+\mathrm{St}^{-1}}}} c_{\rm s},
    \label{turb_vel}
\end{equation}
where $\mathrm{St}$ is the Stokes number corresponding to dust grains of maximum size $a_{\rm max}$, and $c_{\rm s}$ is the sound speed. We note that this form of $u_{\rm turb}$ correctly reproduces a drop in $u_{\rm turb}$ at large $\mathrm{St}$. We also note that defining $\mathrm{St}$ as above can somewhat overestimate its value, and that weighting the stopping time over the dust size distribution in the $[0.2 a_{\rm max}:a_{\rm max}]$ bin could be a better option. The thermal velocity of the Brownian motion is calculated as
\begin{equation}
    u_{\rm th} = \sqrt{ 16 k_{\rm B} T_{\rm d} \over \pi m_{\rm d}},
\end{equation}
where $k_{\rm B}$ is the Boltzmann constant and $T_{\rm d}$ is the dust temperature.

%---------------------------------------------------%

When calculating the volume density of dust, we take into account dust settling by calculating the effective scale height of grown dust  $H_{\rm d}$ via the corresponding gas scale height $H_{\rm g}$, $\alpha$ parameter, and the Stokes number as
\begin{equation}
    H_{\rm d} = H_{\rm g} \sqrt{\frac{\alpha}{\alpha + \mathrm{St}}}.
    \label{eq:dust-scale-height}
\end{equation}
The scale height of gas is determined assuming the local hydrostatic equilibrium in the gravitational field of the central star and the disk \citep{Vorobyov2009}.
The Stokes number is defined as:
\begin{equation}
    {\rm St} = \frac{\Omega_{\rm K}\rho_{\rm s}a_{\rm max}}{\rho_{\rm g}c_{\rm s}},
\end{equation}
where $\Omega_{\rm K}=\sqrt{GM_{\ast}/r^{3}}$ represents the Keplerian angular velocity with the central stellar mass $M_{\ast}$.

%---------------------------------------------------%

Dust growth in our model is limited by collisional fragmentation and drift.  We note that the drift barrier is accounted for self-consistently via the computation of the grown dust dynamics. The fragmentation barrier is taken into account by calculating the characteristic fragmentation size as \citep{Birnstiel2016}:
\begin{equation}
    a_{\rm frag}=\frac{2\Sigma_{\rm g} \mathit{u}_{\rm frag}^2}{3\pi\rho_{\rm s} \alpha c_{\rm s}^2},
\label{eq:afrag}
\end{equation}
where $\mathit{u}_{\rm frag}$ is the fragmentation velocity, namely, a threshold value of the relative velocity of dust particles at which collisions result in fragmentation rather than coagulation. In the current study, we adopt $\mathit{u}_{\rm frag} = 3$\:m\:s$^{-1}$ \citep{Blum2018}. If $a_{\rm max}$ becomes greater than $a_{\rm frag}$, we halt the growth of dust and set $a_{\rm max} = a_{\rm frag}$.

%---------------------------------------------------%

We finally note that the dust radial drift velocity can be distinguished between gradient and advective drift velocities \citep{Birnstiel2016}. The former does not directly depend on the $\alpha$ parameter but is rather determined by the radial gradient of the gas pressure in the disk. The latter relies on the $\alpha$ parameter and reflects the radial advection of dust with gas. In our simulation, since the $\alpha$ parameter is set relatively low, the gradient drift dominates \citep[see also][]{Vorobyov2023}. Therefore, the drift motion of dust grains in our models is determined by the magnitude and radial direction of the gas pressure gradient rather than by the $\alpha$ parameter.

%%%%%%%%%%%%%%%%%%%%%%%%%%%%%%%%%%%%%%%%%%%%
%%% Section 2.3 %%%%
\subsection{Initial conditions}
\label{Sect:initial}

%---- TABLE ----%
\begin{table*}
\center
\caption{\label{Tab:1}The initial parameters of the pre-stellar cloud cores}
\begin{tabular}{c c c c c c c}
\hline 
\hline 
$\Sigma_0$ & $c_{{\rm s},0}$ & $T_0$ & $\rho_0$ & $r_0$ & $\Omega_0$ & $M_{\rm cloud}$ \tabularnewline
g\:cm$^{-2}$ & km\:s$^{-1}$ & K & g\:cm$^{-3}$ & au & km\:s$^{-1}$\:pc$^{-1}$ & $\Msun$ \tabularnewline
\hline
$9.8\times10^{-2}$ & 0.21 & 12 & $3.8\times10^{-18}$ & 1700 & 2.1 & 0.88 \tabularnewline
\hline
\end{tabular}
\center{ \textbf{Notes.} $\Sigma_0$ is the surface density at the plateau, $c_{{\rm s},0}$ is the initial sound speed, $T_0$ is the initial temperature, $\rho_0$ is the initial mass density, $r_0$ is the plateau radius, $\Omega_0$ is the angular velocity at the plateau, and $M_{\rm cloud}$ is the cloud mass.}
\end{table*}

%---------------------------------------------------%

The initial gas surface density and angular velocity profiles of the cloud core we set are derived from an axisymmetric cloud collapse where the angular momentum remains constant \citep{Basu1997}:
\begin{equation}
    \Sigma_{\rm g} = \frac{\Sigma_{0}}{\sqrt{1 + \left( r/r_{0} \right)^2}},
\label{Eq:Sigma0}
\end{equation}
\begin{equation}
    \Omega = 2\Omega_{0}\left( \frac{r_{0}}{r} \right)^2 
    \left[ \sqrt{1 + \left( \frac{r}{r_{0}} \right)^2} - 1 \right].
\label{Eq:Omega0}
\end{equation}
The above profile has a plateau with a uniform surface density extending to the radius $r_{0}$, $\Sigma_{0}$ and $\Omega_{0}$ are the surface density and angular velocity at the plateau, respectively. The plateau radius $r_{0}$ is proportional to the Jeans length
\begin{align}
    r_{0} = A \frac{c_{{\rm s},0}}{\sqrt{\pi G \rho_{0}}}, 
\label{Eq:r0}
\end{align}
where $A$ is a constant parameter, $c_{{\rm s},0}$ is the initial sound speed, $G$ is the gravitational constant, and $\rho_{0}$ is the initial mass density. The initial mass and the plateau surface densities are related as $\Sigma_{0} = r_{0}\rho_{0}$. The constant $A$ is a parameter defining the initial density perturbation. We set $A = \sqrt{1.2}$, with which the ratio of the cloud thermal and gravitational energies is 0.8. We also set the ratio of the rotational and gravitational energies as $7\times10^{-3}$ by adjusting the plateau angular velocity $\Omega_{0}$. 

%---------------------------------------------------%

We calculate a model with one-tenth of solar metallicity (0.1\:$\Zsun$), which corresponds to a dust-to-gas mass ratio of $10^{-3}$. The initial dust-to-gas mass ratio in the gravitationally contracting cloud is assumed to be spatially uniform.

%---------------------------------------------------%

To constrain the free parameters of the initial surface density configuration, namely, $c_{{\rm s},0}$ and $\rho_0$, we make use of a one-zone model developed by \cite{Omukai2005}, which derives the relationship between the central density and temperature in contracting clouds.  To simulate the evolution of the disk formed from a cloud of nearly solar mass, we select a density and temperature at which the Jeans mass approximates solar mass, based on the derived density-temperature relationship of the one-zone model with 0.1\:$\Zsun$. These selections help determine the initial mass density $\rho_{0}$ and the initial sound speed $c_{{\rm s},0}$. The plateau radius is then calculated from Equation\:(\ref{Eq:r0}) using these initial parameters, and the initial gas surface density $\Sigma_0$ is determined as $\Sigma_{0} = r_{0}\rho_{0}$. Table\:\ref{Tab:1} summarizes the values for the initial pre-stellar cloud parameters. Regarding the dust surface density, we assume that it consists solely of small dust grains, and its density is scaled from the gas surface density in accordance with the dust-to-gas mass ratio.

%---------------------------------------------------% 

The pre-stellar cloud defined by Equations~(\ref{Eq:Sigma0})--(\ref{Eq:r0}) has the form of a flattened disk-like configuration in the thin-disk approximation and is initially gravitationally unstable due to a uniform positive density perturbation with amplitude $A=\sqrt{1.2}$. The cloud contracts in the $(r,\:\phi)$ plane, while simultaneously spinning up. The centrifugally balanced protoplanetary disk forms when the in-spiraling material of the contracting cloud hits the centrifugal barrier near the inner computational boundary. This occurs when the centrifugal radius of the infalling gas exceeds the sink cell radius of 10\:au after about twice the free-fall time elapsed from the onset of cloud collapse. Subsequently, the disk gains matter from the infalling parental cloud while losing it via accretion onto a star introduced within the sink cell. We follow the evolution of the disk and the growth of dust for 750\:kyr after the disk formation. We note that all materials from the envelope land on the outer edge of the disk under the thin-disk approximation. This is a reasonable assumption according to \cite{Visser2009}, who calculated gas trajectories in the three-dimensional envelope and found that the bulk of infalling materials accrete on the outer edge of the disk. In the initial stages of collapse, vertical infall of material with low angular momentum, originally located along the rotational axis of the deprojected three-dimensional cloud, may be important, but this effect is neglected in the current version of the code.

%---------------------------------------------------%

The numerical grid contains $1024 \times 1024$ cells, which are logarithmically spaced in the radial direction and linearly in the azimuthal one, providing sub-au resolution up to a distance of $\approx 150$\:au. Since our model disk becomes gravitationally unstable in its early evolution stage and is also prone to gravitational fragmentation, the numerical grid should resolve the local Jeans length $R_{\rm J}\simeq c_{\rm s}^2 / (G \Sigma_{\rm g})$ by at least four numerical cells to avoid spurious fragmentation \citep{Truelove1998,Vorobyov2013}. Fragments usually condense out of the densest sections of spiral arms. The typical surface densities and temperatures in spiral arms do not exceed 100\:g\:cm$^{-2}$ and 100\:K in the early, prone-to-fragmentation evolution period. The corresponding Jeans length is $R_{\rm J} \approx 20$\:au. The radial and azimuthal grid resolution at $r = 100$\:au is $\approx 0.66$\:au and at $r=200$\:au is $\approx 1.3$\:au.  Most fragments form at these distances and, thus, the Jeans length is well resolved.

%%%%%%%%%%%%%%%%%%%%%%%%%%%%%%%%%%%%%%%%%%%%
%%%%%%%%%%%%%%%%%%%%%%%%%%%%%%%%%%%%%%%%%%%%
%%% SECTION 3 %%%%
\section{Disk evolution}
\label{Sec:DiskEvolution}

%---------------------------------------------------%

We present our simulation results in Sections\:\ref{Sec:EarlyPhase} and \ref{Sec:LatePhase}. In Section\:\ref{Sec:EarlyPhase}, we describe the early-phase disk evolution, corresponding to the first 300\:kyr after disk formation, during which disk fragmentation and dust growth from micron to millimeter sizes occur. Then, in Section\:\ref{Sec:LatePhase}, we focus on the late-phase evolution, which stretches from 300\:kyr to 750\:kyr, characterized by the cessation of disk fragmentation and the emergence of multiple dust rings.

%%%%%%%%%%%%%%%%%%%%%%%%%%%%%%%%%%%%%%%%%%%%
%%% SECTION 3.1 %%%%
\subsection{Early phase: dust size growth}
\label{Sec:EarlyPhase}

%---- FIGURE ----%
\begin{figure*}
 \begin{center}
 \begin{tabular}{c} 
  {\includegraphics[width=1.5\columnwidth]{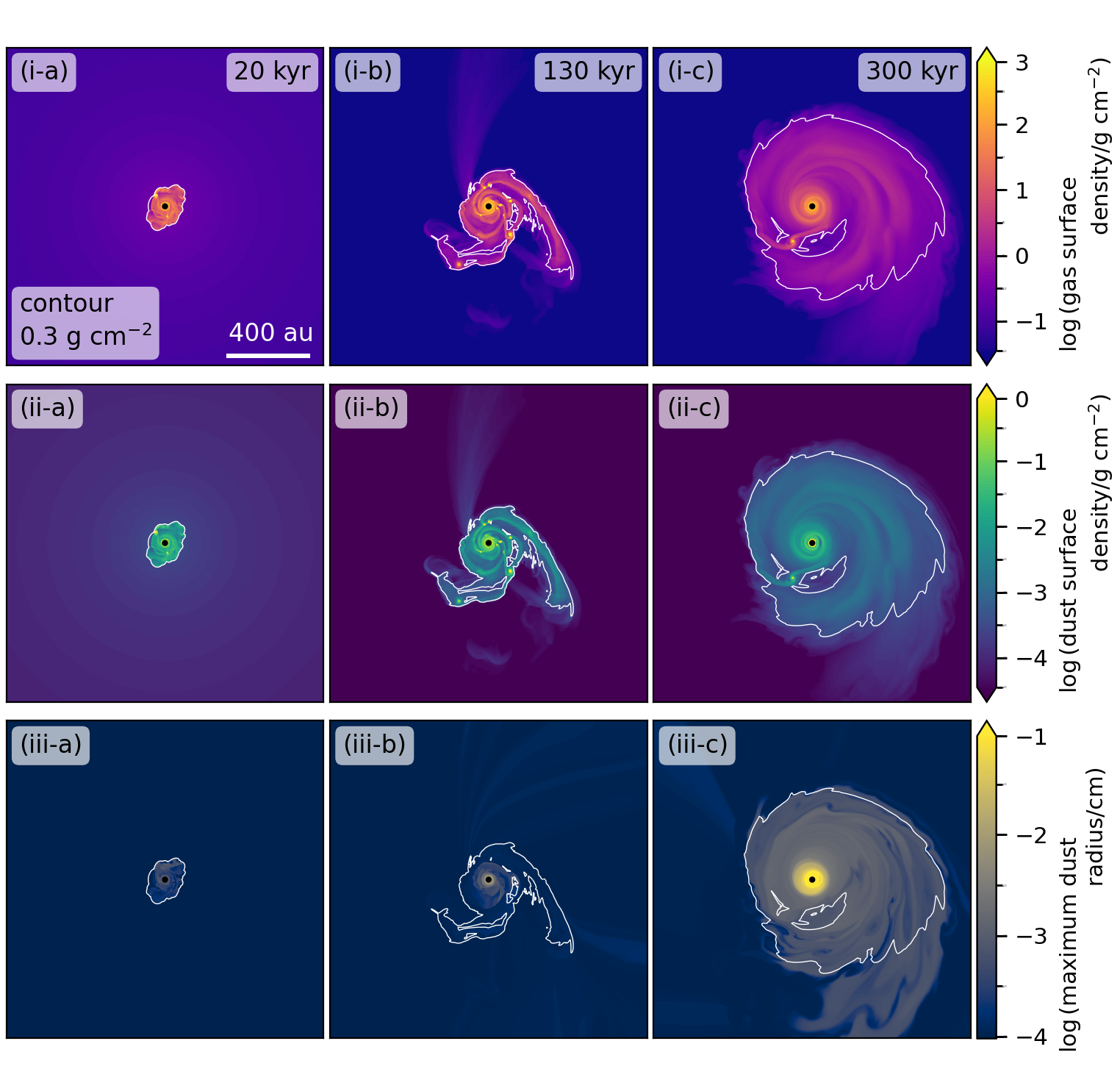}}
 \end{tabular}
 \caption{
 Disk evolution during the first 300\:kyr (early phase). The top, middle, and bottom rows show the 2D distributions of the surface density of gas, that of dust, and the maximum dust size at three different epochs, 20, 130, and 300\:kyr after the disk formation. The dust density is the sum of grown and small dust densities. In each panel, the white contour line delineates an isosurface density of 0.3\:g\:cm$^{-2}$, indicating the approximate location of the outer edge of the disk.
  }
 \label{Fig:2D_EarlyPhase}
 \end{center}
\end{figure*}

%---------------------------------------------------%

Figure\:\ref{Fig:2D_EarlyPhase} shows the temporal evolution of the disk from its formation to 300\:kyr. The top row panels depict the spatial distribution of the gas surface density, demonstrating the progressive expansion of the disk's size. The white contour lines represent isosurface density of 0.3\:g\:cm$^{-2}$, which are used as an approximate indicator of the disk's outer boundary.

%---------------------------------------------------%

Panel\:i-b reveals the presence of non-axisymmetric substructures, such as spiral arms, within the disk, along with $\sim10$ clumps. This indicates that the disk is gravitationally unstable, and the disk fragmentation occurs. The disk fragmentation is most pronounced at this epoch. Subsequently, the disk fragmentation gradually diminishes, and by 300\:kyr (Panel\:i-c), only a single clump remains within the disk. This clump continues to orbit for an additional 100\:kyr. The other clumps migrate inward due to gravitational interactions with spiral arms and with each other, eventually accreting onto the central star or dispersing due to tidal effects.

%---------------------------------------------------%

The density distributions of dust shown in Panels\:ii-a, ii-b, and ii-c closely resemble those of gas. Spiral arms and clumps appear at the same locations as in the gas distribution. At the scale of approximately 100\:au shown in the panels, the decoupling between gas and dust grains is insufficient to produce noticeably different distributions. In later stages, however, differences between these distributions become more apparent (see Section\:\ref{Sec:LatePhase}). For example, at 410\:kyr, the dust-to-gas mass ratio around 100\:au deviates from the initial value ($10^{-3}$) by a factor of 2--3 due to the radial drift of grown dust grains with sizes of 0.1--1\:mm.

%---------------------------------------------------%

Panel\:iii-a shows that the largest dust size throughout the disk is $\sim10^{-4}$\:cm, with only minor signs of dust growth. By 130\:kyr (Panel\:iii-b), grown dust grains with $\sim10^{-3}$\:cm are observed only in the innermost region of the disk ($\sim$10--20\:au), although the maximum dust size across the rest of the disk remains $10^{-4}$\:cm. At 300\:kyr (Panel\:iii-c), the spatial distribution of the maximum dust size undergoes significant changes, with dust growth observed across the entire disk. In the outer regions, the maximum dust size eventually reaches $\sim10^{-3}$ to $10^{-2}$\:cm, and in the inner regions ($\lesssim 30$\:au), it is larger than 1\:cm.

%---- FIGURE ----%
\begin{figure*}
 \begin{center}
 \begin{tabular}{c} 
  {\includegraphics[width=1.95\columnwidth]{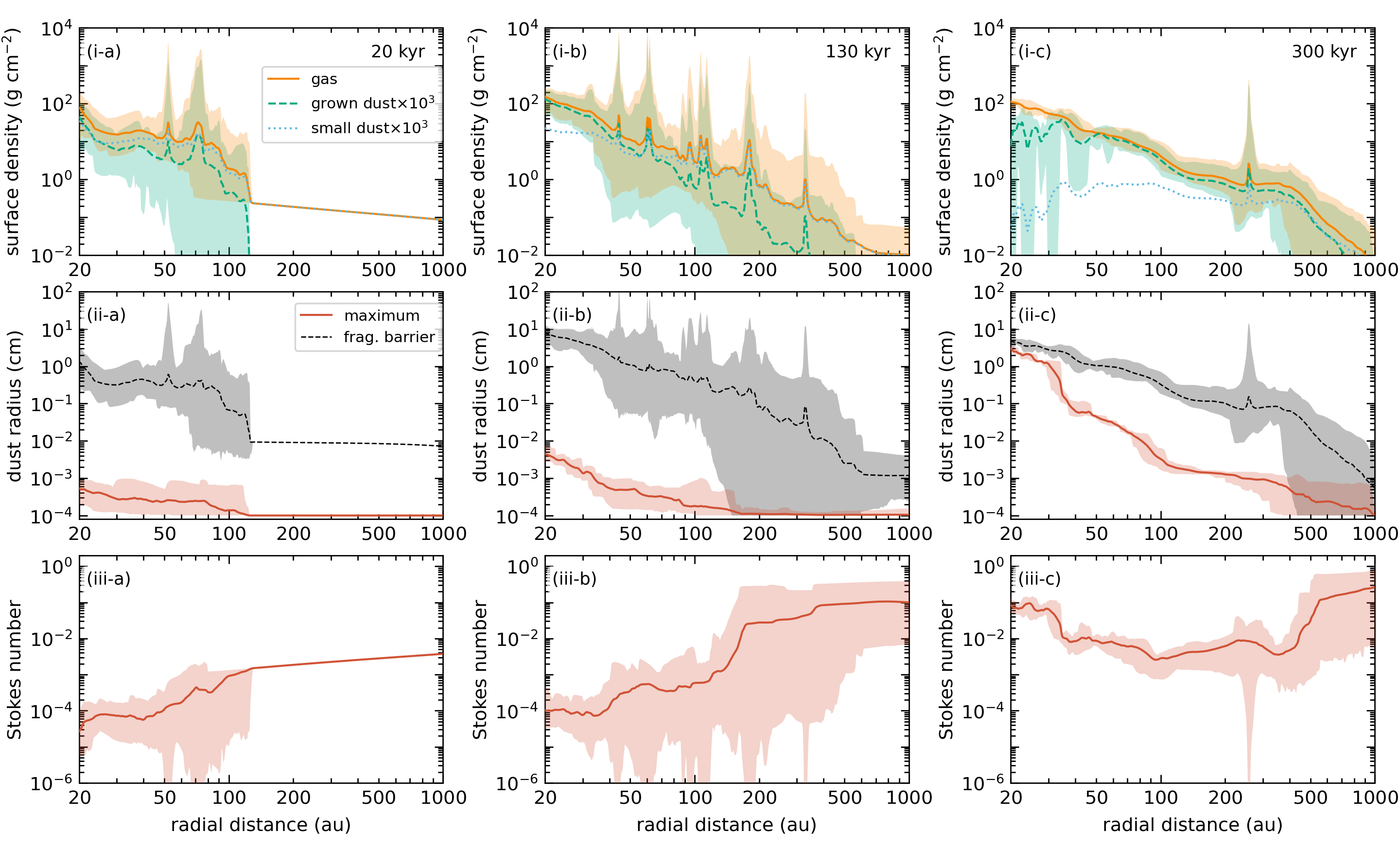}}
 \end{tabular}
 \caption{
 Radial profiles of disk structure and dust properties in the early phase. Each row shows the radial profiles of the surface densities (top), the maximum dust size (middle) and the Stokes number (bottom) at the same epochs as in Figure\:\ref{Fig:2D_EarlyPhase}. The lines provide the azimuthally averaged values, and the shaded layers show the ranges of variation of these values at a given radius, with the upper and lower boundaries corresponding to the maximum and minimum values. In the top panels, the lines and colors indicate different components: gas (orange solid), grown dust (green dashed), and small dust (blue dotted). For clarity, the shaded layers are shown only for the gas and grown dust components. The densities of the grown and small dust components are multiplied by 1000 to make comparisons easier. In the middle panels, the lines depict the maximum dust size (red solid line) and the fragmentation barrier (black dashed line). 
 }
 \label{Fig:Radial_EarlyPhase}
 \end{center}
\end{figure*}

%---------------------------------------------------%

Figure\:\ref{Fig:Radial_EarlyPhase} illustrates the radial distributions of various properties of gas and dust grains. In the top row, the surface densities of gas, grown dust, and small dust are displayed. Panel\:i-a shows a gradual increase in the gas density toward the center from 1000 to 130\:au, indicating the presence of the infalling envelope. A significant increase in gas density is observed at 130\:au, defining the outer boundary of the disk. There are two peaks at 50 and 70\:au within the disk, corresponding to clumps formed due to disk fragmentation. In Panel\:i-b, multiple peaks are observed because, at this epoch, disk fragmentation is most intense, resulting in the formation of many clumps. In Panel\:i-c, the density variation in the azimuthal direction decreases within 200\:au (see the orange shaded area) as the disk becomes marginally stabilized.

%---------------------------------------------------%

In Panel\:i-a, since we assume that all dust grains initially belong to the small component, dust grains within the accreting envelope are almost entirely small. Grown dust appears within the disk because the density becomes sufficiently high for growth via dust mutual collisions, marking the onset of dust growth. By 130\:kyr (Panel\:i-b), the density of grown dust exceeds that of small dust in the inner disk region ($< 50$\:au). By 300\:kyr (Panel\:i-c), the density of grown dust is entirely higher than that of small dust throughout the disk. Since the grown dust component becomes dominant in the disk, decoupled motion between gas and dust is expected thereafter.

%---------------------------------------------------%

The middle row of Figure\:\ref{Fig:Radial_EarlyPhase} shows the radial distributions of the maximum dust size. At 30\:kyr, the dust size in all regions is less than $10^{-3}$\:cm (Panel\:ii-a). The growth of dust size progresses from the inner regions of the disk and, by 130\:kyr, reaches sub-mm sizes in the innermost areas (Panel\:ii-b). Beyond 50\:au, however, the dust size remains below $10^{-3}$\:cm. By 300\:kyr, the dust size reaches mm scales at $\sim50$\:au and grows to approximately sub-mm scales at $\sim100$\:au (Panel\:ii-c). These results align with analytical estimates of the timescale for collisional dust growth from micron to mm sizes, given by \citet{Birnstiel2016}
\begin{align}
  t_{\rm growth} &\simeq \left( \frac{\Sigma_{\rm d}}{\Sigma_{\rm g}} \right)^{-1} \frac{1}{\Omega_{\rm K}} \notag \\
  &\sim 200 \left( \frac{\Sigma_{\rm d}/\Sigma_{\rm g}}{10^{-3}} \right)^{-1} \left( \frac{M_{\ast}}{0.5\:\Msun} \right)^{-1/2}  \notag \\
  &\times\left( \frac{r}{100\:{\rm au}} \right)^{3/2}\:{\rm kyr},
  \label{Eq:t_growth}
\end{align}
where we assume coagulation growth between dust grains of similar size, consistent with the dust growth scheme used in our simulations. Additionally, at 300\:kyr, the dust size in the inner regions (20--30\:au) approaches the fragmentation barrier, as shown by the black dashed line. In contrast, in other regions, the dust size is well below the fragmentation barrier, with the drift barrier limiting the dust size. The observed pattern of dust growth is also consistent with theoretical expectations, with the maximum dust size in the inner and outer disk regions being limited by fragmentation and drift barriers, respectively \citep{Birnstiel2016}.

%---------------------------------------------------%

The Stokes number is directly related to the growth of dust grains. As shown in Panels\:iii-a and iii-b of Figure\:\ref{Fig:Radial_EarlyPhase}, before significant dust growth, the average Stokes number is below $10^{-3}$ between 20\:au and 100\:au, indicating small values throughout the disk. This indicates that the motion of dust grains within the disk is almost coupled with that of the gas in the early phase. By 300\:kyr, as dust grains grow, the Stokes number increases accordingly (Panel\:iii-c). At 20\:au, where the dust size reaches cm scales, the Stokes number is $\sim$0.1. Across the disk, the Stokes number at 300\:kyr reaches 0.01, which is one to two orders of magnitude larger than its value at 130\:kyr. Dust grains with Stokes numbers of $\gg 10^{-3}$ are expected to decouple from gas motion on disk evolution timescales, a process necessary for dust enhancements in local pressure maxima. Our long-term simulation enables unified tracking of dust growth and the subsequent decoupling motion in a subsolar-metallicity disk.

%%%%%%%%%%%%%%%%%%%%%%%%%%%%%%%%%%%%%%%%%%%%
%%% SECTION 3.2 %%%%
\subsection{Late phase: multiple dust ring formation}
\label{Sec:LatePhase}

%---------------------------------------------------%

We continue to follow the evolution of the disk that has undergone intense fragmentation, as described in Section\:\ref{Sec:EarlyPhase}. We obtain results with a long-term evolution of 750\:kyr, being close to the typical lifetime of gas disks, which is impossible for fully three-dimensional disk evolution simulations with a comparable numerical resolution.

%---- FIGURE ----%
\begin{figure}
 \begin{center}
 \begin{tabular}{c} 
  {\includegraphics[width=0.9\columnwidth]{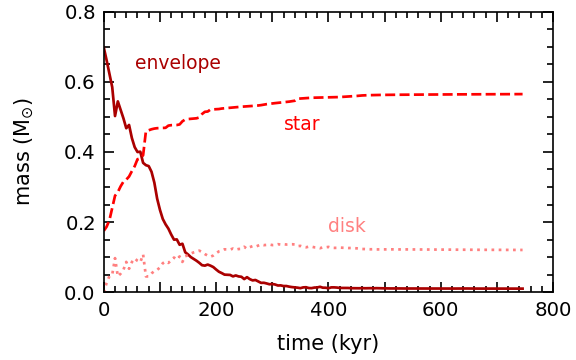}}
 \end{tabular}
 \caption{
Evolution of the gas mass distribution up to the end of the fiducial run. The lines represent three different components, the envelope (solid), the central star (dashed), and the disk (dotted).
 }
 \label{Fig:Mass}
 \end{center}
\end{figure}

%---------------------------------------------------%

The disk mass varies over time. Figure\:\ref{Fig:Mass} shows the temporal evolution of the masses of the disk, the central star, and the envelope. To measure these masses, we take azimuthal averages of gas density and velocity, identifying the disk's radius as the location where the density exceeds 0.1\:g\:cm$^{-2}$ and the azimuthal velocity surpasses the radial velocity. Regions within this radius are designated as the disk, while those outside are defined as the envelope. Initially, the density threshold does not align with the disk's radius immediately after disk formation, but it aligns well after about 100\:kyr (see Figure\:\ref{Fig:2D_EarlyPhase}). The velocity condition is crucial for correctly determining the disk radius in the early phase. Figure\:\ref{Fig:Mass} shows that, immediately after disk formation, the majority of the gas mass resides in the envelope (0.7\:$\Msun$). By 300\:kyr, the envelope rapidly loses mass, while the masses of the central star and disk increase to 0.55\:$\Msun$ and 0.15\:$\Msun$, respectively. The slower increase in disk mass compared to the star is attributed to efficient gas transfer from the disk to the star, driven by self-gravitational instability. After 400\:kyr, the envelope mass approaches zero, indicating the depletion of gas accreting from the envelope to the disk. Moreover, the mass growth of both the star and the disk plateaus.

%---- FIGURE ----%
\begin{figure*}
 \begin{center}
 \begin{tabular}{c} 
  {\includegraphics[width=1.99\columnwidth]{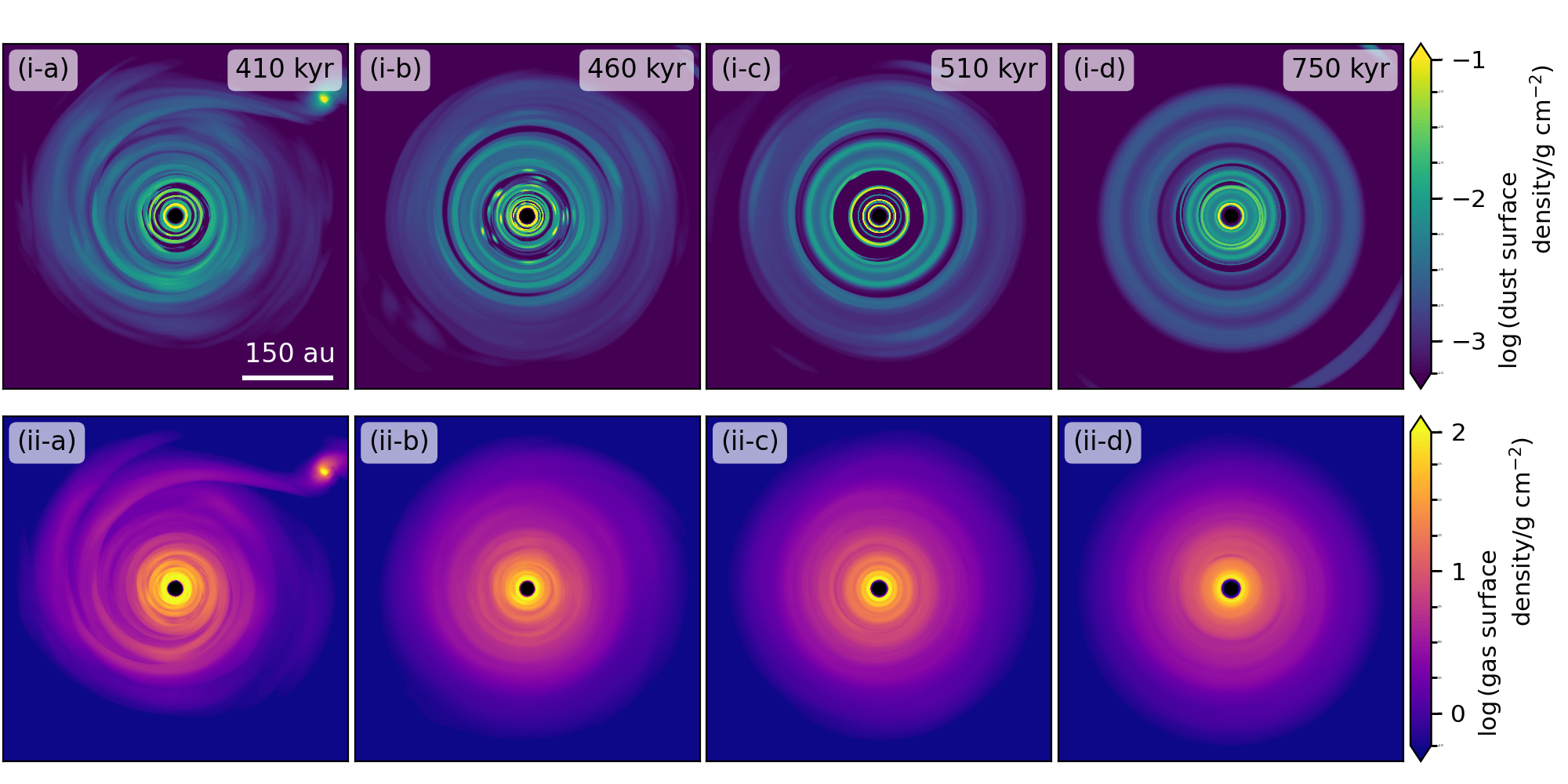}}
 \end{tabular}
 \caption{
 Multiple dust ring formation in the late phase. Each column displays the spatial distributions of the surface densities of dust (upper) and gas (lower) at four different periods, 410, 460, 510, and 750\:kyr from left to right.
 }
 \label{Fig:DustRing}
 \end{center}
\end{figure*}

%---------------------------------------------------%

Figure\:\ref{Fig:DustRing} depicts the formation of multiple dust rings within 300\:au, and the upper and lower panels show the dust and gas surface density structures, respectively. In the upper right corner of Panel\:i-a, there is a clump which is the same as the one observed in Panel\:i-c of Figure\:\ref{Fig:2D_EarlyPhase}. The dust-to-gas mass ratio inside this clump remains close to the initial value of $10^{-3}$. No new clumps appear because the values of Toomre's Q parameter \citep{Toomre1964} range from 2 to 5 across the entire disk, indicating that the disk is already stable against fragmentation due to the depletion of mass infall from the envelope \citep[e.g.][]{Vorobyov2005}. A spiral arm extends from the clump, tightly winding toward the center. The clump dissipates shortly after this epoch, likely due to tidal forces from the central star and Keplerian shear in the surrounding gas \footnote{At 410\:kyr, the Roche surface density of the clump is estimated to be $\sim$12\:g\:cm$^{-2}$. Using the clump identification scheme of \citet{Matsukoba2022}, its mass and radius are $\sim$$9.3\times10^{-4}$\:$\Msun$ and 5.4\:au, respectively, yielding an average surface density of $\sim$90\:g\:cm$^{-2}$. Since this value exceeds the Roche density, the clump is not expected to be tidally disrupted by the central star. Nevertheless, our simulation shows that the clump dissipates. A more precise assessment of its fate would require higher spatial resolution capable of resolving the contraction phase of the clump, which is beyond the scope of the present study.}. Following its dissipation, non-axisymmetric substructures become less prominent, and axisymmetric substructures are gradually highlighted (Panels\:i-b and i-c). Panel i-c clearly shows multiple axisymmetric dust rings and gaps. These dust rings persist for a long time and are still observable at the end of this simulation at 750\:kyr (Panel\:i-d).

%---------------------------------------------------%

We note that after disk fragmentation a clump is in general expected to undergo quasi-static contraction and evolve into a protoplanet with a sub-au radius \citep{Helled2011}. As the clump becomes more compact, the tidal effect and Keplerian shear would be reduced, and the fate of the clump might be changed. In our case, the clump at 410\:kyr is located at $\sim$300\:au, where the grid resolution of 2.0\:au is insufficient to capture this effect.   Incorporating the effect of such contraction requires a higher spatial resolution or adaptive mesh refinement, which is beyond the scope of present study.

%---------------------------------------------------%

The dust properties in the late phase, such as dust size and Stokes number, do not differ significantly from those at $\sim$300\:kyr (see Figure\:\ref{Fig:Radial_EarlyPhase}). Within 200\:au, where the dust rings are located, the dust size ranges from 0.01 to 1\:cm, and the Stokes number lies between 0.01 and 0.1. Although the dust size at the dust ring radii becomes larger than in the surrounding regions, the difference is modest, typically by a factor of about two.

%---------------------------------------------------%

In contrast to the dust, the lower panels of Figure\:\ref{Fig:DustRing} show that the radial variation of gas density is modest, with faint peaks and gaps. In Panel\:ii-a, the spiral arm extending from the clump is distinctly visible, similar to Panel\:i-a. After the dissipation of the clump, in Panels\:ii-b and ii-c, the gas substructures flatten over time due to the effects of turbulent viscosity.

%---- FIGURE ----%
\begin{figure}
 \begin{center}
 \begin{tabular}{c} 
  {\includegraphics[width=0.95\columnwidth]{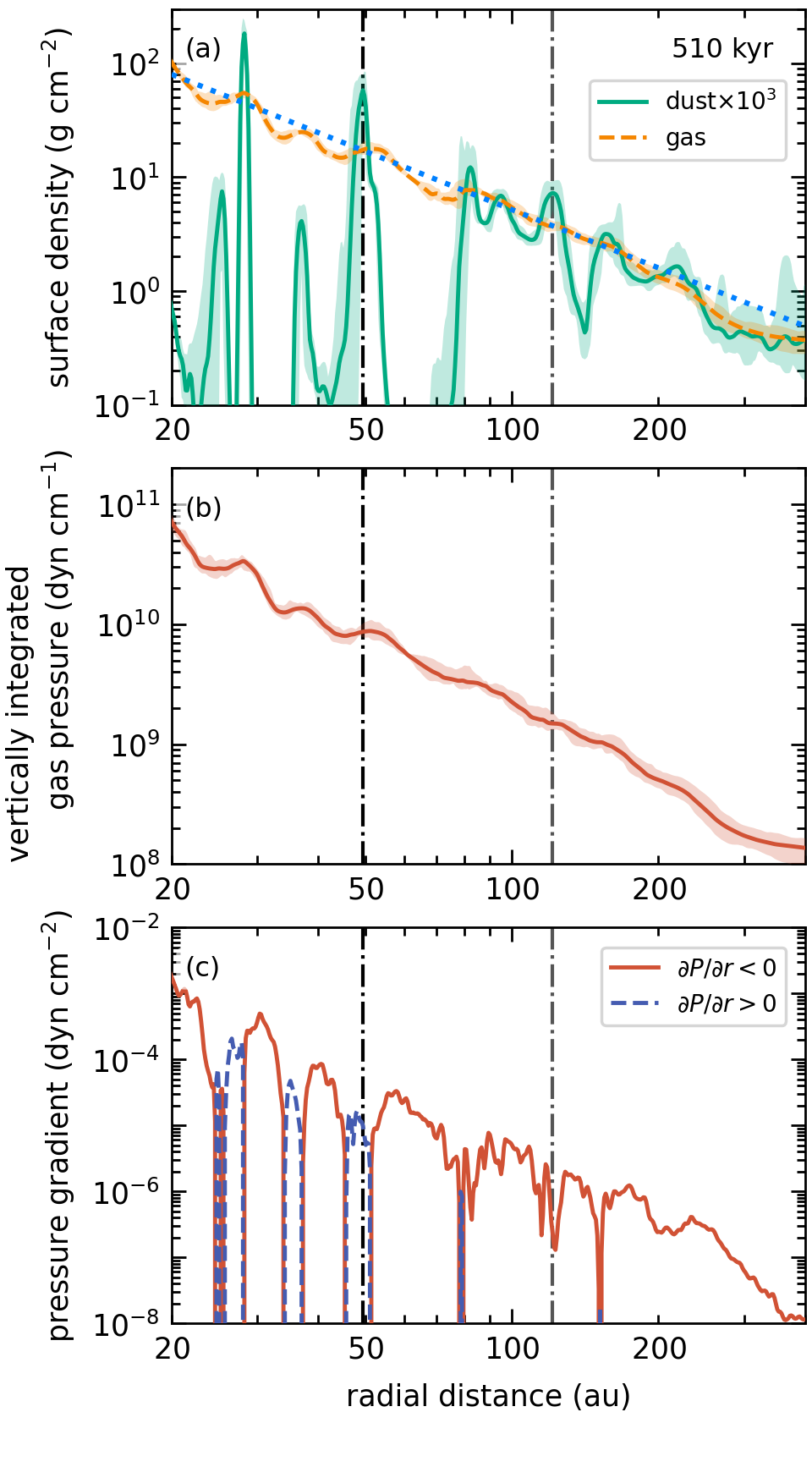}}
 \end{tabular}
 \caption{
 Radial distributions of (a) the dust and gas surface densities, (b) the vertically integrated gas pressure, and (c) its radial gradient at 510\:kyr. The lines show the azimuthally averaged values, and the shaded regions represent the range between the maximum and minimum values. In Panel (a), the colors depict different components: dust (green) and gas (orange). The dust density is multiplied by 1000 to facilitate easier comparison. In Panel (c), the red solid line depicts negative gradients, and the blue dashed line indicates positive gradients. The dash-dotted vertical lines mark the positions of two examples of dust density peaks, at 50\:au (black) and 120\:au (gray), among several observed. In Panel (a), the blue dotted line represents the fitting function for the gas density, given by $80\left( r/20\:{\rm au} \right)^{-1.7}$\:g\:cm$^{-2}$, which is used in Section \ref{Sec:NumericalExperiments}.
 }
 \label{Fig:RadialDensPress}
 \end{center}
\end{figure}

%---------------------------------------------------%

Here, we detail the radial structure of the disk after the formation of multiple dust rings and gaps. Figure\:\ref{Fig:RadialDensPress}a displays the radial distributions of the dust and gas surface densities at 510\:kyr. Several peaks of the dust density are located within 300\:au, corresponding to dust rings. The four dust rings within 60\:au exhibit more pronounced radial density variations compared to the surrounding gap regions, and the densities of the dust rings from 70\:au to 300\:au are amplified a few times compared to the gap regions. The gas density has less pronounced variations than that of the dust density. 

%---------------------------------------------------%

Figures\:\ref{Fig:RadialDensPress}b and \ref{Fig:RadialDensPress}c show the radial distributions of vertically integrated gas pressure and its radial gradient, respectively. The gas pressure exhibits fluctuations similar to those of the gas density, with several peaks. These fluctuations in gas pressure result in a radius-dependent radial pressure gradient. Generally, in protoplanetary disks, both density and temperature increase toward the inner regions, leading to a monotonic increase in gas pressure inward, and the radial gradient of gas pressure is negative at all radii. However, because the disk in this simulation has variations in the radial distribution of gas density, there are radii where the radial pressure gradient becomes positive or locally minimal.

%---------------------------------------------------%

The radial gradient of gas pressure affects the direction and speed of dust drift motion \citep{Weidenschilling1977}. At radii where the radial pressure gradient is negative, dust grains receive headwinds from the gas and drift inward. Conversely, at radii where the radial pressure gradient is positive, dust grains receive tailwinds from the gas and drift outward. Therefore, at radii where the radial gradient switches from negative to positive when moving inward from the outer disk, the direction of dust grains’ drift also changes. As a result, the dust grains tend to pile up at that location. An example of dust grains accumulating at a certain radius due to the switching radial gradient is the dust ring at 50\:au as indicated by the vertical black dash-dotted line in Figure\:\ref{Fig:RadialDensPress}.

%---------------------------------------------------%

Some dust rings also form at positions where the radial gradient does not switch from negative to positive, such as the dust ring at 120\:au (gray vertical dash-dotted line in Figure\:\ref{Fig:RadialDensPress}). At these positions, the radial gradient is negative, but its magnitude is locally minimal. The speed of dust drift motion in the radial direction depends on the magnitude of the radial gradient \citep{Birnstiel2016}. Therefore, at locations where the radial gradient is locally minimal, the speed of dust drift motion slows down. This causes a ``traffic jam'' of dust grains, leading to an enhanced dust density at that position compared to its surroundings \citep{DrazkowskaAlibert2017}.

%---------------------------------------------------%

The distance between dust rings (gap widths) within 100\:au is on the order of 10\:au. For example, the rings are located at $\simeq$ 50\:au and $\simeq$ 80\:au, and their separation is $\simeq$ 30\:au. The dust ring formation mechanism presented in this study originates from gravitational instability, for which the most unstable wavelength is given by 
\begin{align}
    \lambda_{\rm max} = \frac{2 c_{\rm s}^2}{G\Sigma_{\rm g}}.
\label{Eq:lambda_max}
\end{align}
Based on the gas scale height 
\begin{align}
    H_{\rm g} = \frac{c_{\rm s}}{\Omega_{\rm K}},
\end{align}
and the Toomre's Q parameter 
\begin{align}
    \mathcal{Q}_{\rm T} = \frac{c_{\rm s}\Omega}{\pi G\Sigma_{\rm g}},
\end{align}
Equation\:\eqref{Eq:lambda_max} is modified as
\begin{align}
    \lambda_{\rm max} = 2\pi \mathcal{Q}_{\rm T}H_{\rm g}.
\end{align}
Since the Toomre's Q parameter during the period of dust ring formation ranges from 2 to 5, the most unstable wavelength is approximately one order of magnitude larger than the gas scale height. Indeed, at 50\:au, the gas scale height is 3.1\:au, which matches well with the observed dust ring separation, indicating consistency between the dust ring spacing and the most unstable wavelength of gravitational instability.

%%%%%%%%%%%%%%%%%%%%%%%%%%%%%%%%%%%%%%%%%%%%
%%%%%%%%%%%%%%%%%%%%%%%%%%%%%%%%%%%%%%%%%%%%
%%% SECTION 4 %%%%
\section{Varying levels of turbulent viscosity}
\label{Sec:NumericalExperiments}

%---------------------------------------------------%

Here we conduct numerical experiments to investigate the conditions under which the axisymmetric dust rings appear in our simulation (Section\:\ref{Sec:LatePhase}). Whereas we have used $\alpha = 10^{-4}$ for the fiducial run, here we consider two additional experimental runs where the turbulent viscosity is stronger and negligible, i.e., $\alpha = 10^{-3}$ and 0 (inviscid gas), respectively. We emphasize that gravitational torques become the sole mass transport mechanism in the latter case. We initiate these runs from a snapshot in the fiducial run. Specifically, we use the saved data at the epoch of 410\:kyr, which includes gas and dust values (e.g., densities, velocities, temperatures, etc.). We only change the value of the $\alpha$ parameter for these runs.

%---- FIGURE ----%
\begin{figure*}
 \begin{center}
 \begin{tabular}{c} 
  {\includegraphics[width=1.9\columnwidth]{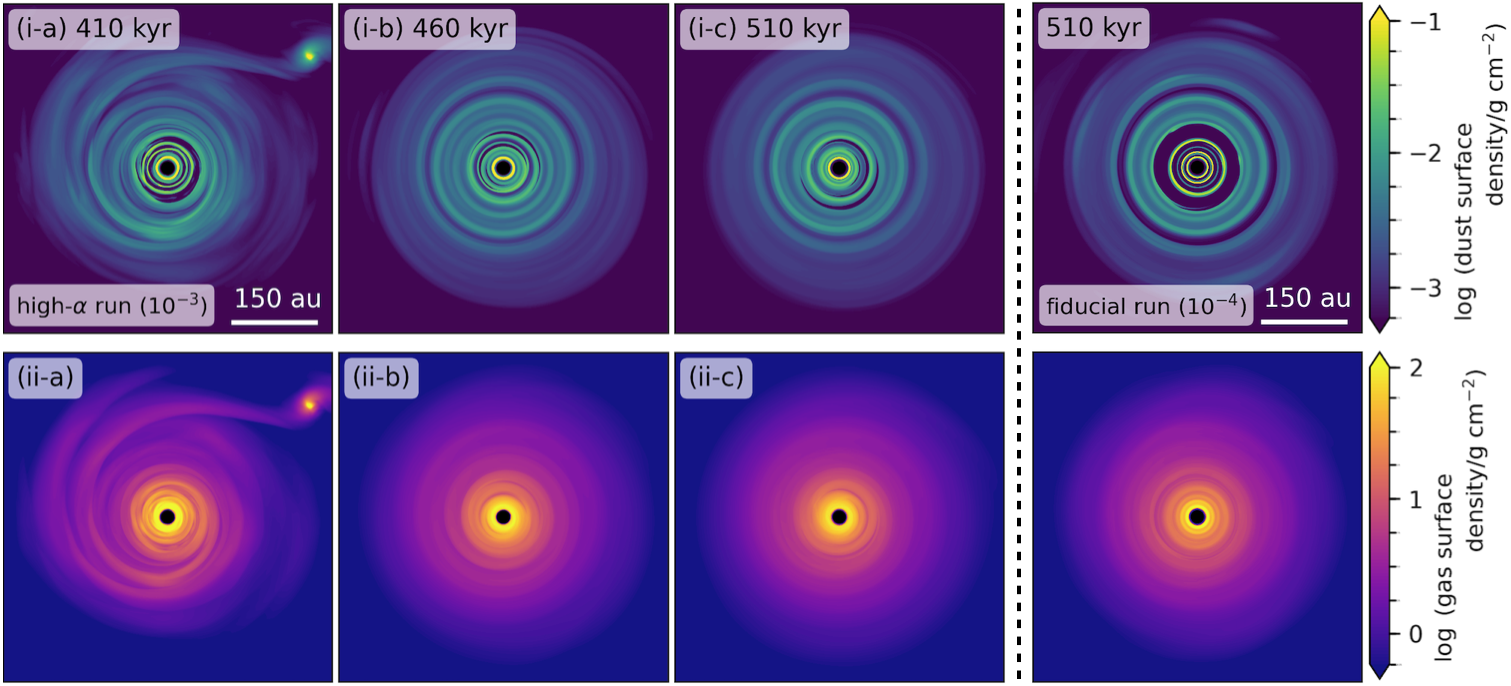}}
 \end{tabular}
 \caption{
 Spatial distributions of the dust and gas surface densities in the run with high turbulent viscosity of $\alpha=10^{-3}$ at three different epochs, 410, 460, and 510\:kyr, shown in the three columns on the left. In the rightmost column, the spatial distributions in the fiducial run with the viscosity of $\alpha = 10^{-4}$ are displayed.
 }
 \label{Fig:Alpha-3}
 \end{center}
\end{figure*}

%---- FIGURE ----%
\begin{figure}
 \begin{center}
 \begin{tabular}{c} 
  {\includegraphics[width=0.95\columnwidth]{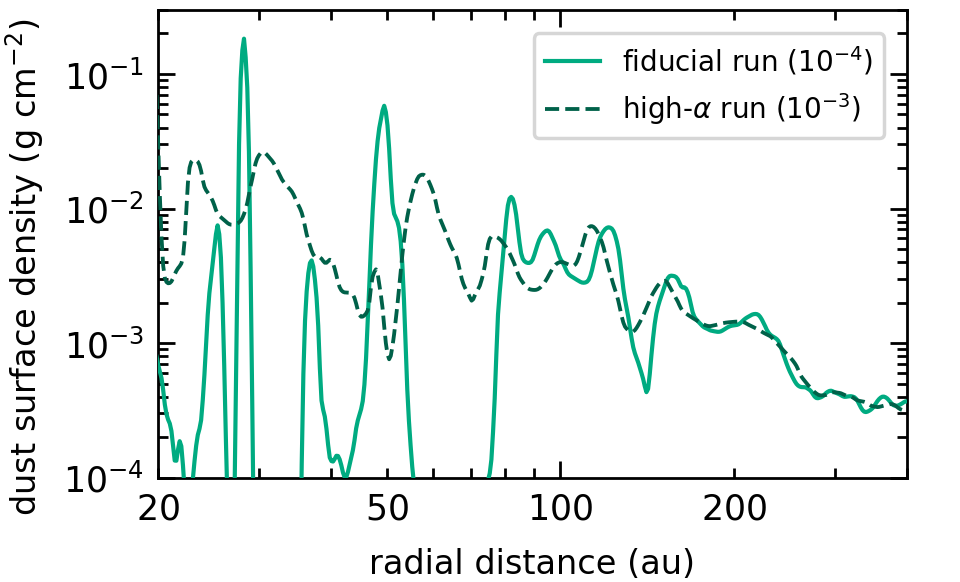}}
 \end{tabular}
 \caption{
 Radial profiles of the dust surface density at 510\:kyr in the fiducial (solid) and high-$\alpha$-viscosity (dashed) runs. 
 }
 \label{Fig:RadialDustDensity}
 \end{center}
\end{figure}

%---------------------------------------------------%

Figure\:\ref{Fig:Alpha-3} shows the disk evolution in the run with the higher turbulent viscosity, $\alpha=10^{-3}$. The formation of multiple dust rings is also observed in this run (Panels\:i-b and i-c), similar to the fiducial run. Panel\:i-b shows that the axisymmetric dust rings form by 460\:kyr, indicating an earlier formation compared to the fiducial run. However, we see that the dust gaps appear more filled in than in the fiducial run. 

%---------------------------------------------------%

Figure\:\ref{Fig:RadialDustDensity} compares the radial distributions of the dust surface density in the runs with $\alpha=10^{-4}$ (fiducial run) and $10^{-3}$. In the fiducial run, the dust gaps within 50\:au have densities three orders of magnitude lower than the peak values of the neighboring dust rings (solid line). In contrast, with $\alpha=10^{-3}$, the dust gaps are shallower, with densities only about an order of magnitude lower than the peak values of the neighboring dust rings (dashed line). This suggests that in disks with $\alpha=10^{-3}$, the dust density forms a flatter structure, and dust gaps are less pronounced. Dust rings may disappear in disks with even stronger turbulent viscosities ($\alpha > 10^{-3}$), although this conjecture has to be verified with numerical experiments.

%---- FIGURE ----%
\begin{figure*}
 \begin{center}
 \begin{tabular}{c} 
  {\includegraphics[width=1.9\columnwidth]{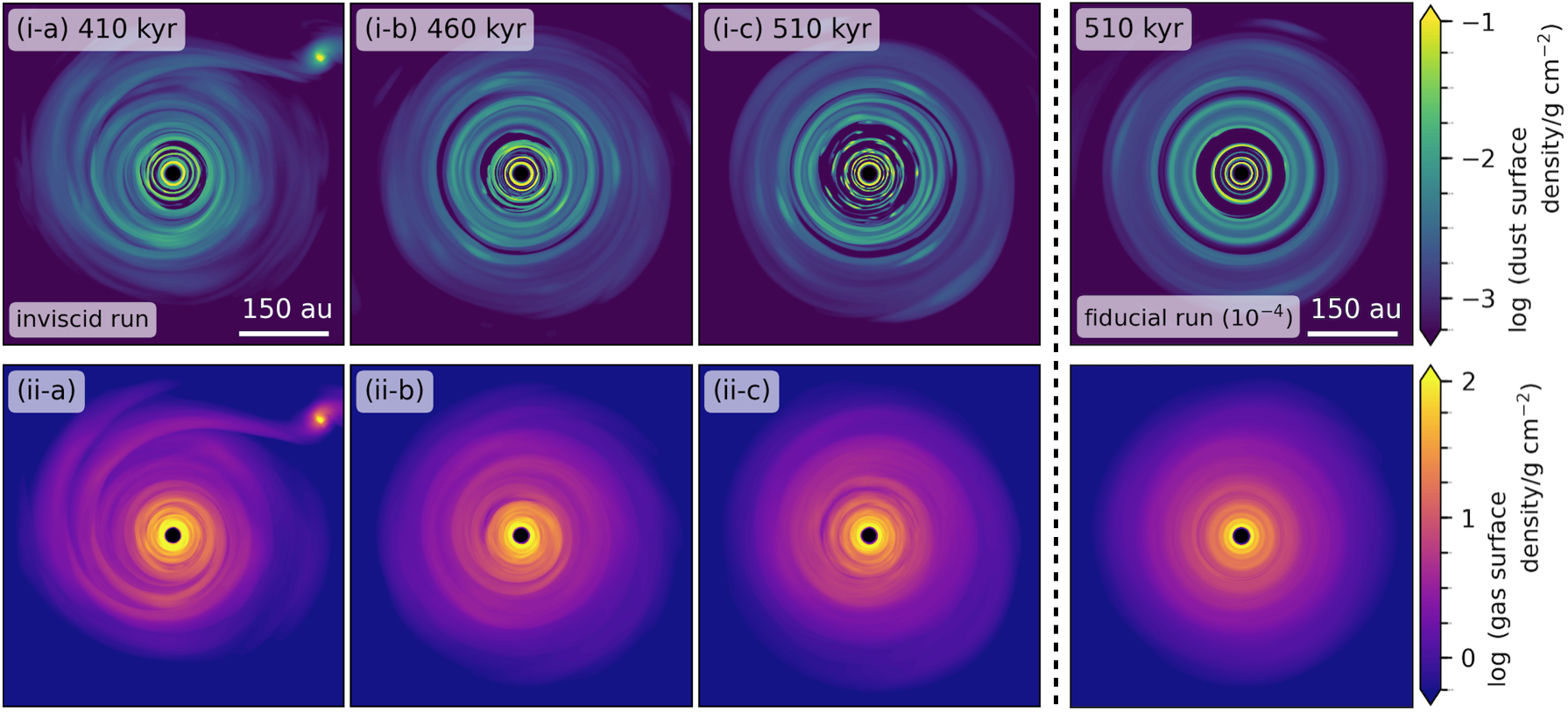}}
 \end{tabular}
 \caption{
 Same as Figure\:\ref{Fig:Alpha-3}, but for the inviscid run.
 }
 \label{Fig:inviscid}
 \end{center}
\end{figure*}

%---------------------------------------------------%

Figure\:\ref{Fig:inviscid} shows the disk evolution in the inviscid run, where the effects of the turbulent viscosity are neglected by setting $\alpha=0$. In this run, an obvious axisymmetric dust substructure does not form. In Panels\:i-b and i-c, there are variations in dust density in the azimuthal direction. This result indicates that the formation of dust rings requires finite levels of the turbulent viscosity. This numerical experiment also demonstrates that the magnitude of numerical viscosity in FEOSAD is much smaller than an equivalent of $\alpha=10^{-4}$, otherwise we would not have seen much difference with the fiducial run. Low numerical viscosity is ensured by the third-order accurate advection scheme used in the code \citep[see also Appendix in][]{Vorobyov2007}.

%---------------------------------------------------%

We note that, if the dust distribution in this run is observed with high angular resolution, the non-axisymmetric substructures may appear as crescent-shaped features. In observations with limited resolution, such non-axisymmetric structures can be smoothed out and appear axisymmetric.

%---- FIGURE ----%
\begin{figure}
 \begin{center}
 \begin{tabular}{c} 
  {\includegraphics[width=0.87\columnwidth]{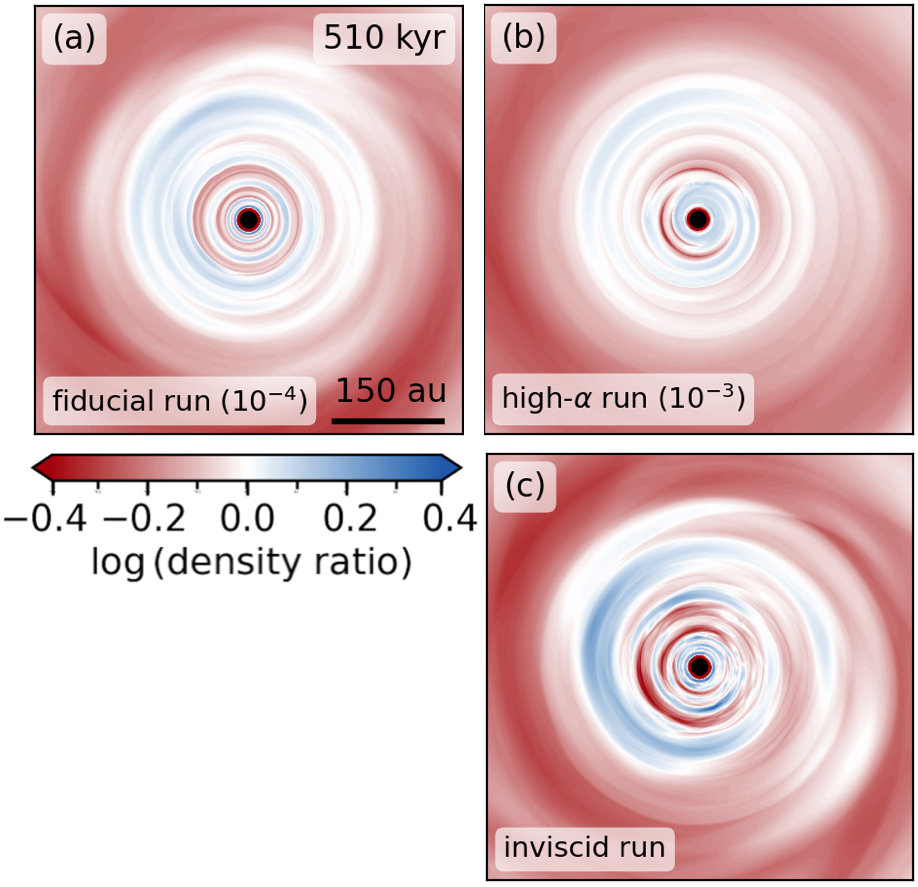}}  
 \end{tabular}
 \caption{
 Gas surface density ratio between the simulation results and the fitting function shown in Figure\:\ref{Fig:RadialDensPress}a, given by $80\left( r/20\:{\rm au} \right)^{-1.7}$\:g\:cm$^{-2}$. The panels represent the snapshot at the same epoch of 510\:kyr in the three different simulations: (a) the fiducial, (b) high-alpha-viscosity, and (c) inviscid runs. In each panel, the blue regions indicate areas of relatively higher surface density compared to their surroundings, while the red regions indicate areas of relatively lower density.
 }
 \label{Fig:DensityRatio}
 \end{center}
\end{figure}

%---------------------------------------------------%

To examine the impact of variations in turbulent viscosity on the gas substructures, we compare the gas densities in the above three simulations to a reference density in the top panel of Figure\:\ref{Fig:DensityRatio}. The reference density is derived from the radial distribution of gas density at 510\:kyr (Figure\:\ref{Fig:RadialDensPress}) as $80\left( r/20\:{\rm au} \right)^{-1.7}$\:g\:cm$^{-2}$. Panel\:c shows that the density ratio in the inviscid run exhibits non-axisymmetric deviations that are more pronounced than those in the two runs with $\alpha=10^{-4}$ and $10^{-3}$ (Panels a and b). These deviations correspond to the spiral arm structures associated with the clump and persist without being sheared out.

%---------------------------------------------------%

All panels (a, b, and c) show crescent-shaped structures. In Panel b, a feature appears at a radial distance of about 50\:au, where the density is noticeably lower than in the surrounding disk. This structure emerges at $\sim$440\:kyr and persists at nearly the same radial distance until at least 510\:kyr, while being associated with a higher temperature than its surroundings. The feature is more pronounced in the $\alpha=10^{-3}$ run (Panel b) and in the inviscid run (Panel c), but is less evident in the $\alpha=10^{-4}$ model. We note that in the latter model the dust rings are most pronounced. The physical origin of the crescent-shaped structures, however, remains unclear and should be addressed in future work.

%---------------------------------------------------%

The experimental runs mentioned above suggest that an appropriate magnitude of the turbulent viscosity is essential for creating the axisymmetric multiple dust rings. There should exist a lower limit of the $\alpha$ parameter necessary to convert the original tightly wound spiral arms into clearly axisymmetric rings, as well as an upper limit to avoid excessively smoothing out the gas density peaks. Although our simulations show that some levels of turbulence around $\alpha \sim 10^{-4}$ are necessary, determining the exact viscosity range will require further detailed studies.

%%%%%%%%%%%%%%%%%%%%%%%%%%%%%%%%%%%%%%%%%%%%
%%%%%%%%%%%%%%%%%%%%%%%%%%%%%%%%%%%%%%%%%%%%
%%% SECTION 5 %%%%
\section{Summary and discussion}
\label{Sec:Summary}

We have investigated the formation of multiple dust rings and gaps in a protoplanetary disk with metallicity 0.1\:$\Zsun$ by performing a two-dimensional hydrodynamic simulation. The simulation follows the long-term evolution of the disk up to 750\:kyr after its formation. This duration is sufficiently long for dust grains to grow to sizes on the order of millimeters in the subsolar-metallicity disk. Thanks to this long-term simulation, we have identified a new mechanism for dust ring formation, which occurs as the disk stabilizes from a self-gravitationally unstable state in its early stages.

%---------------------------------------------------%

In the early stages, the disk grows in mass and size through mass accretion from its surrounding envelope and becomes gravitationally unstable. The gravitationally unstable disk undergoes fragmentation, with the most intense fragmentation observed around 100\:kyr after its formation. By 300\:kyr, disk fragmentation subsides, leaving tightly wound spiral arms in the disk.

%---------------------------------------------------%

In the late stages, gas accretion from the envelope diminishes, and the disk transitions from a gravitationally unstable state toward a more stable state. As this transition progresses, the non-axisymmetric spiral arms evolve into axisymmetric structures. These axisymmetric gas structures generate several gas pressure bumps across tens to hundreds of au. Dust grains are trapped in these pressure bumps through their drift motion. The multiple dust rings form at $\sim$510\:kyr and persist for $\sim$240\:kyr, until the end of the simulation at 750\:kyr. This marks a considerably prolonged phase since the formation of the circumstellar disk and represents an epoch before which dust size growth occurs even at 0.1\:$\Zsun$, enabling the gas and dust to move independently of each other.

%---------------------------------------------------%

During the formation of dust rings, a finite level of turbulent viscosity is required to facilitate the non-axisymmetric spiral arms to evolve into axisymmetric structures. If the $\alpha$-parameter representing the viscosity is larger than $10^{-3}$, the density gaps in the gas will fill in, preventing the formation of dust rings and gaps. Conversely, if it is smaller than $10^{-4}$, the transition to axisymmetric structures will not occur. More studies are needed to determine the appropriate viscosity range.

%-- DISCUSSION --%

%---- FIGURE ----%
\begin{figure}
 \begin{center}
 \begin{tabular}{c} 
  {\includegraphics[width=0.95\columnwidth]{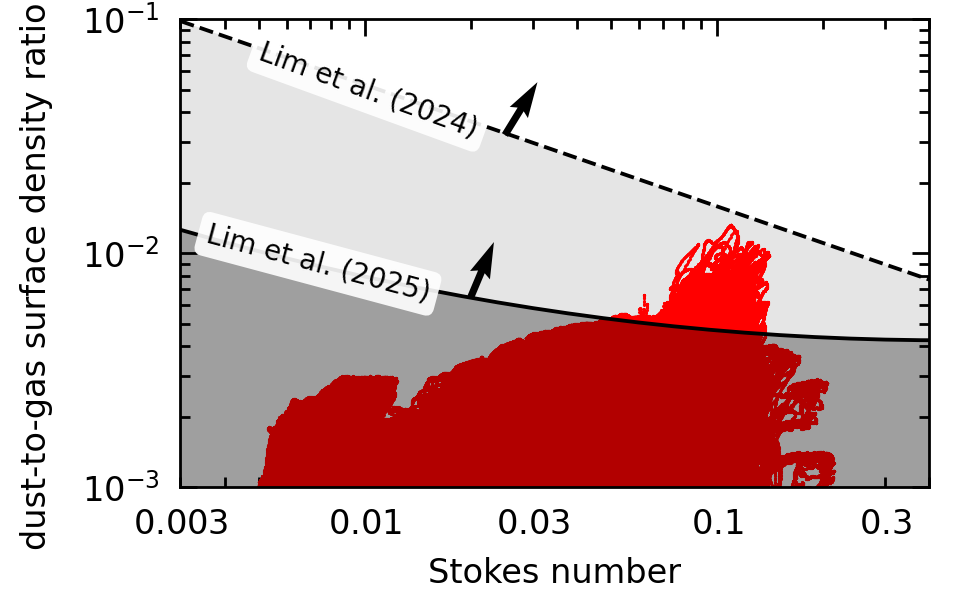}}
 \end{tabular}
 \caption{
 Conditions for planetesimal formation via the streaming instability. Red dots represent the Stokes number and dust-to-gas surface density ratio for each grid cell, which spans from 30 to 300\:au, sampled every 2\:kyr from 510\:kyr to the end of the simulation at 750\:kyr. The solid and dashed lines indicate thresholds for strong clumping due to the streaming instability, based on criteria from \cite{Lim2025} and \cite{Lim2024}, respectively, with the latter corresponding to the turbulent $\alpha=10^{-4}$. Regions above either of the thresholds would undergo strong clumping conducive to planetesimal formation.
 }
 \label{Fig:StreamingInst}
 \end{center}
\end{figure}

%---------------------------------------------------%

Streaming instability is one of the leading mechanisms for forming planetesimals from dust within dust rings \citep[e.g.][]{Youdin2005, Johansen2007, Carrera2015, Yang2017, Li2021}. Several numerical studies have examined the clumping of mm- and cm-sized dust via streaming instability. These studies provide the dust-to-gas mass ratios required for streaming instability to occur as a function of the Stokes number. We examined if conditions in the dust rings are conducive to the development of streaming instability by considering the latest criteria provided by \citet{Lim2024} and \citet{Lim2025}. The former is applicable to the disk conditions when some sort of external turbulent force is present in the disk, such as provided by the magnetorotational instability, and is more relevant to our work. 

%---------------------------------------------------%

Figure\:\ref{Fig:StreamingInst} indicates that conditions in the dust rings fulfill the \citet{Lim2025} criterion with a large margin, but do not meet the more stringent \citet{Lim2024} criterion. Typically, the dust rings located at radii of 50--80\:au correspond to the region where the \citet{Lim2025} criterion is most likely satisfied. Recent work by \citet{Tominaga2025} has shown that considering dust coagulation during streaming instability can relax the \citet{Lim2024} criterion. This coagulation-assisted effect suggests that, even in cases with external turbulence as in our simulations, dust rings may still undergo strong clumping. We also note that dust particles grow much above 1.0\:mm, see Figure\:\ref{Fig:Radial_EarlyPhase}, and thus overcome the threshold for the streaming instability to develop in pressure traps according to \citet{Carrera2022}. Furthermore, considering the settling of dust grains and gas removal processes, such as photoevaporation \citep[e.g.][]{Hollenbach2000,  Owen2012, Gorti2015, Carrera2017, Nakatani2018} and magnetohydrodynamic disk winds \citep[e.g.][]{Suzuki2010, Suzuki2016, Bai2016, Tabone2022}, the local dust-to-gas mass ratio in the midplane can be further enhanced, potentially creating even more favorable conditions for streaming instability and planetesimal formation. Therefore, we express a cautious optimism that our proposed mechanism can lead to planetesimal formation in protoplanetary disks with metallicity as low as 0.1\:$Z_\odot$.

%---------------------------------------------------%

In this study, we have focused exclusively on the subsolar-metallicity disk and discussed the new mechanism for dust ring formation. The necessary conditions for this mechanism include tightly wound spiral arms created by self-gravitational instability, dust grains large enough to decouple from gas, and moderate viscosity. Provided these factors are in place, dust ring formation will be possible not only in subsolar-metallicity disks but also in those with solar metallicity. The observed values of turbulent viscosity range from $10^{-4}$ to $10^{-3}$ \citep[e.g.][]{Pinte2016, Dullemond2018, Trapman2020}, which align with the viscosity values in our numerical experiments where dust ring formation has been confirmed. However, observations of nearby Class I disks do not show non-axisymmetric structures \citep{Ohashi2023}, raising questions about the presence of tightly wound spiral arms in early-stage disks. These observational results are under debate regarding the effects of viewing angle and surrounding gas. More detailed observations are needed.

%---------------------------------------------------%

Our numerical simulation has tracked the evolution of a subsolar-metallicity disk from its formation up to 750\:kyr. The lifetime of such disks is observationally estimated to be $\sim$1\:Myr \citep{Yasui2010, Yasui2016, Yasui2021, Guarcello2021}, shorter than the 3--6\:Myr typically inferred for disks in solar-metallicity regions \citep{Haisch2001, Hernandez2007, Meyer2007, Mamajek2009, Ribas2014, Richert2018}. Thus, our simulation period already approaches the expected lifetime of low-metallicity disks. Within this period, multiple dust rings are present from 510\:kyr until the end of the run at 750\:kyr, demonstrating a survival time of at least $\sim$240\:kyr. This longevity reflects a disk that was initially massive enough to become gravitationally unstable and subsequently stabilized, allowing dust rings to survive for a substantial fraction of the disk’s lifetime. By contrast, less massive low-metallicity disks that remained stable throughout would likely disperse more rapidly and may not form long-lived rings via this mechanism. Solar-metallicity disks, on the other hand, tend to persist for several Myr, although current observational lifetimes do not constrain how long such disks may have spent in a self-gravitating phase in their early evolution. A more systematic exploration of the long-term evolution of disks with different metallicities and initial masses will be essential for assessing how generally the dust-ring formation process we identified operates. In particular, simulations including gas-dissipation processes such as photoevaporation \citep[e.g.][]{Nakatani2018, Nakatani2018a} are required to determine the diversity of dust-ring lifetimes and their implications for planet formation.

%%%%%%%%%%%%%%%%%%%%%%%%%%%%%%%%%%%%%%%%%%%%
%%%%%%%%%%%%%%%   ACKNOWLEDGMENTS   %%%%%%%%%%%%%%%
\begin{acknowledgments}
The authors would like to thank Sanemichi Takahashi, Ryosuke Tominaga, Hidekazu Tanaka, and Shu-ichiro Inutsuka for valuable discussions and insightful comments.  
We are also grateful to the anonymous reviewer for constructive feedback that helped to improve the manuscript. This research has been made possible thanks to the support by Grants-in-Aid for Scientific Research (TH: 19KK0353, 21H00041) from the Japan Society for the Promotion of Science. E.I.V acknowledges support by the Ministry of Science and Higher Education of the Russian Federation (state contract FEUZ-2023-0019). T.H. appreciates financial support from ISHIZUE 2024 of Kyoto University and Kyoto University Foundation. Numerical computations were carried out on Cray XC50 at the Center for Computational Astrophysics (CfCA) of the National Astronomical Observatory of Japan and the Vienna Scientific Cluster (VSC-4). 
\end{acknowledgments}

%%%%%%%%%%%%%%%%%%%%%%%%%%%%%%%%%%%%%%%%%%%%
%%%%%%%%%%%%%%%%   CONTRIBUTION   %%%%%%%%%%%%%%%%%%
\begin{contribution}
R.M. conducted and analyzed the numerical simulation data and was responsible for writing and submitting the manuscript.
E.I.V. conducted the numerical simulations, obtained the funding, and edited the manuscript.
T.H. also obtained the funding and edited the manuscript.
\end{contribution}

%---------------------------------------------------%

\software{\texttt{matplotlib} \citep{Hunter2007}}

%%%%%%%%%%%%%%%%%%%%%%%%%%%%%%%%%%%%%%%%%%
%%%%%%%%%%%%%%%%% APPENDICES %%%%%%%%%%%%%%%%%
\appendix

\section{Effects of the inner boundary condition}
\label{Sec:Boundary}

As was noted in \citet{VorobyovAkimkin2018} and in references therein \citep[e.g.,][]{Zhu2012}, the standard one-way open boundary condition, which allows matter to flow through the disk-sink interface toward the star but prohibits the flow in the opposite direction, can lead to a strong drop in the density near the sink. This drop can indeed be expected at the physical disk inner edge, where the disk is truncated by the interaction with the stellar magnetosphere, but the problem is that this region is difficult to reach in multidimensional disk simulations and the radius of the sink cell is usually much larger than the disk truncation radius.
In this case, the drop in density is usually caused by the lack of compensating flow across the one-way boundary when waves are excited by a planet or in strongly gravitationally unstable disks (as in our case).

\begin{figure}
 \begin{center}
 \begin{tabular}{c} 
  {\includegraphics[width=0.95\columnwidth]{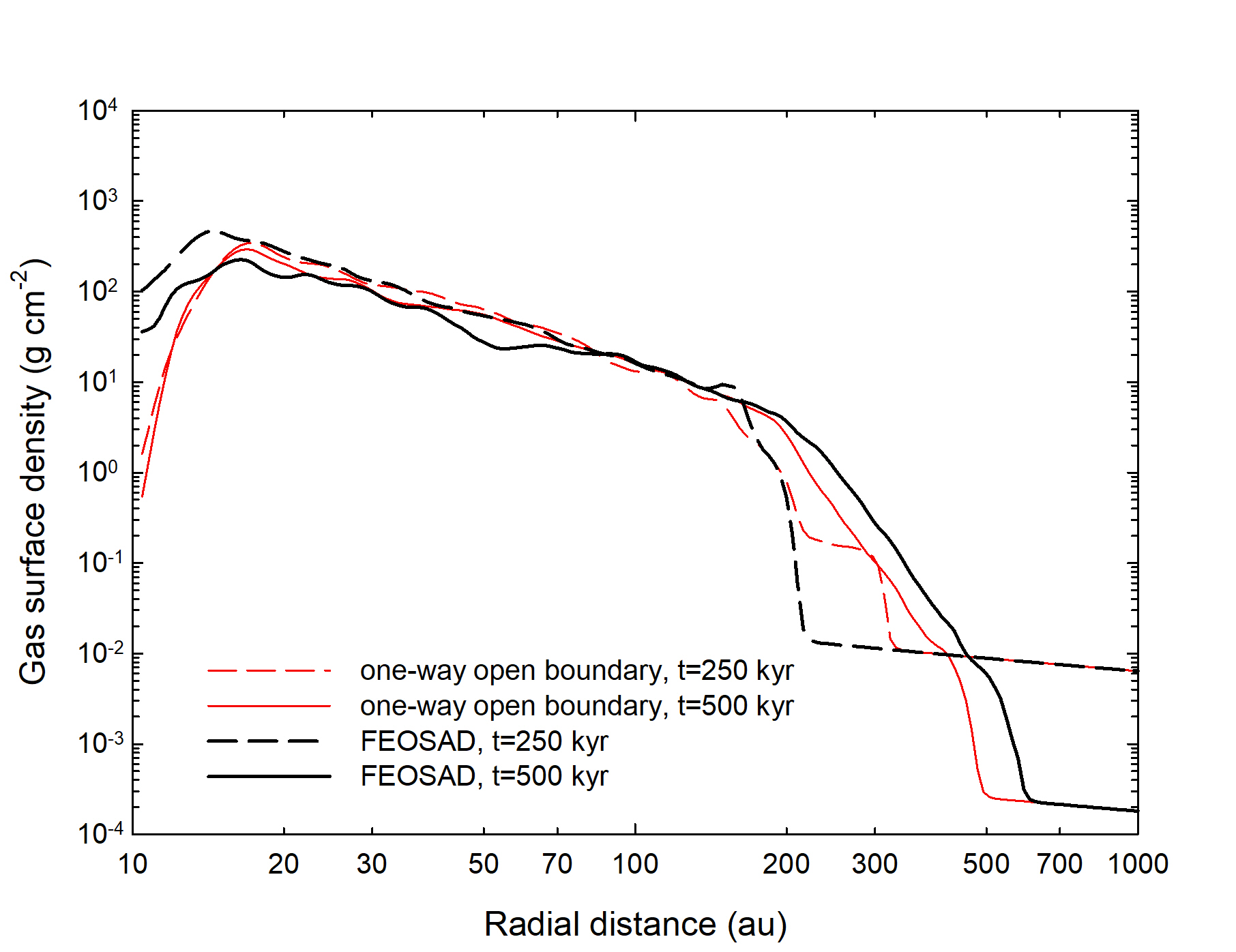}}
 \end{tabular}
 \caption{
 Radial gas surface density distributions in the test models with different inner boundary conditions as indicated in the legend.
 }
 \label{Fig:boundary}
 \end{center}
\end{figure}

To mitigate this problem, \citet{VorobyovAkimkin2018} and \citet{VorobyovSkliarevskii2019} have designed a boundary condition that significantly reduces the undesired drop in density  by constructing a smart sink cell. In this model, matter is allowed to travel both ways across the sink-disk interface, thus providing a compensating flow of matter from the sink to the disk. The details of this boundary condition can be found in the aforementioned papers. Here, we provide the results of a numerical test with both boundary conditions considered. Figure\:\ref{Fig:boundary} presents the gas density radial profiles at $t=$250\:kyr and 500\:kyr after the onset of simulations. Clearly, the one-way open boundary condition yields a much deeper drop near the sink cell (10\:au) than the standard FEOSAD boundary condition. Nevertheless, the inner several astronomical units may still be affected by the boundary condition and we therefore cut them out in our analysis.

%---- FIGURE ----%

%%%%%%%%%%%%%%%%%%%%%%%%%%%%%%%%%%%%%%%%%%
%%%%%%%%%%%%%%%%% REFERENCES %%%%%%%%%%%%%%%%%
\bibliographystyle{aasjournalv7}

\end{document}